\documentclass[oneside,reqno]{amsart}
\usepackage{epsfig}
\usepackage{amsmath,amssymb}
\usepackage{cite}

\newcommand{\bmt}{\begin{pmatrix}}
\newcommand{\emt}{\end{pmatrix}}
\newcommand{\vpu}[1]{^{\vphantom{#1}}}
\newcommand{\myStrut}[1]{\parbox{0.11 pt}{\rule{0 ex}{#1 ex}}}
\newcommand{\te}[2]{\parbox[b]{#1 cm}{ \centering #2}}
\DeclareMathOperator{\SL}{SL}
\DeclareMathOperator{\ESL}{ESL}
\DeclareMathOperator{\Tr}{Tr}
\DeclareMathOperator{\Det}{Det}

\begin{document}
\begin{titlepage}
\begin{center}
\bfseries  Systems of Imprimitivity for the Clifford Group
\end{center}
\vspace{1 cm}
\begin{center} D.M.~Appleby
\\
\emph{Perimeter Institute for Theoretical Physics,  Waterloo, Ontario N2L 2Y5, Canada}
 \end{center}

\vspace{0.2 cm}

\begin{center}
Ingemar Bengtsson\\
 \emph{Stockholms Universitet, AlbaNova,
  Fysikum, S-106 91
  Stockholm, Sweden}
 \end{center}

\vspace{0.2 cm }

\begin{center}
Stephen Brierley
\\
\emph{ Heilbronn Institute for Mathematical Research, Department of Mathematics, University of
  Bristol, Bristol
  BS8 1TW, UK}
\end{center}

\vspace{0.2 cm }

\begin{center}
\AA{}sa Ericsson
\\
\emph{ Matematiska institutionen, Link\"opings Universitet, S-581 83 Link\"oping, Sweden}
\end{center}

\vspace{0.2 cm }

\begin{center}
Markus Grassl
\\
\emph{Centre for Quantum Technologies, National
  University of Singapore, Singapore~117543}
\end{center}

\vspace{0.2 cm }

\begin{center}
Jan-\AA{}ke Larsson
\\
\emph{ Institutionen f\"or Systemteknik,
  Link\"opings Universitet, S-581 83 Link\"oping, Sweden}
\end{center}

\vspace{1 cm}

\begin{center} \textbf{Abstract}

\vspace{0.1cm}

\parbox{12 cm }{It is known  that if the dimension is a
  perfect square the Clifford group can be represented by monomial
  matrices.  Another way of expressing this result is to say that when
  the dimension is a perfect square the standard representation of the
  Clifford group has a system of imprimitivity consisting of  one
  dimensional subspaces.  We   generalize this result to the case of
  an arbitrary dimension.  Let $k$ be the square-free part of the
  dimension.  Then we  show that the standard representation of the
  Clifford group has a system of imprimitivity consisting of
  $k$-dimensional subspaces. To illustrate the use of this result we
  apply it to the calculation of SIC-POVMs (symmetric informationally
  complete positive operator valued measures), constructing exact
  solutions in dimensions 8 (hand-calculation) as well as 12 and 28
  (machine-calculation).}
\vspace{0.35 cm}
\parbox{12 cm }{}
\end{center}
\end{titlepage}

\begin{allowdisplaybreaks}

\section{Introduction}
The Clifford group\footnote{We should remark that there are
  several versions of the Clifford
  group~\cite{Hostens,Gross,Vourdas,Appleby09a}.  In this paper the focus will be on the version defined to be the normalizer of a single copy of the non-Galoisian
  version of the Weyl-Heisenberg group.  However, we believe our
  results are likely to be relevant to the other versions.}
plays a major role in many areas of quantum information.  When
expressed in terms of the standard basis many of the Clifford
unitaries are Hadamard matrices, with the property that every matrix
element has the same absolute value $=1/\sqrt{N}$ (where $N$ is the
dimension).  It is therefore somewhat remarkable that, in the special
case when the dimension is a perfect square, a simple change of basis
will cause every Clifford unitary to become what could be regarded as
the opposite of a Hadamard matrix, namely a monomial matrix, with only
one non-zero element in each row and each column~\cite{monomial}.  In
the following we will generalize this result to arbitrary dimensions.
Given any dimension $N$ there exist unique integers $k,n$ such that
$k$ is square-free (\emph{i.e.}\ $k$ is not divisible by the square of
any prime number) and $N=k n^2$.  We will show that the Clifford group
admits a representation entirely in terms of matrices with the
property that, when written out in block form, each row and each
column contains exactly one non-zero $k\times k$ block.  We refer to
such matrices as $k$-nomial.  It will be seen that the result proved
in Ref.~\cite{monomial} is simply the special case of this for $k=1$.

This result can be re-phrased.  Let $\mathcal{H}$ be $N$-dimensional
Hilbert space.  Then our result is that $\mathcal{H}$ can be written
as a direct sum
\begin{equation}\label{eq:decomposition}
\mathcal{H} = \bigoplus_{r,s=0}^{n-1} \mathcal{H}_{r,s}
\end{equation}
where the subspaces $\mathcal{H}_{r,s}$ are all $k$-dimensional and
where, for every Clifford unitary $U$ and every pair $r,s$
\begin{equation}\label{eq:ImprimSystem}
U \mathcal{H}_{r,s} = \mathcal{H}_{r',s'}
\end{equation}
for some $r',s'$.  Such a set of subspaces is called a system of
imprimitivity~\cite{Hall,Curtis}.  Systems of imprimitivity play an
important role in the theory of group representations as they can be
used to construct an induced representation for the whole group
starting from a representation of a subgroup.  No such application is
in question here since we already have a representation for the group.
There are, however, other applications, as we will see.

Many interesting physical properties are revealed by a deeper understanding
of the relevant group structures and properties. The Clifford group in its various forms is a
good example of this, having applications to fault tolerant quantum computing~\cite{gottesman+99, gottesman97}, measurement based quantum computing~\cite{raussendorf+01, raussendorf+03}, describing the boundary between classically simulable and universal quantum computing~\cite{gottesman98, nielsen+chuang}, quantum non-locality~\cite{Fivel},
mutually unbiased
bases (or MUBs)~\cite{Wootters87,Cormick,Gross,Wootters07,Appleby07b,Appleby09a,Kern,Godsil}, and
symmetric informationally complete measurements (commonly referred to
as SIC-POVMs or, as here, simply
SICs)~\cite{Hoggar,monomial,Godsil,Appleby09a,Zauner,Renes04,Grassl04,Appleby05,Grassl05,Klappenecker,Flammia,Appleby07a,Appleby07b,Kibler,Grassl08,Khatirinejad,Bengtsson08,Fickus,Appleby09b,Scott10,Zhu10a,Zhu10b,Fillipov,Appleby11,Bengtsson11a,Appleby12}.
Although the result proved in this paper is potentially relevant to
all these areas we will illustrate its use in connection with the SIC
existence problem, as that is the route by which we were led to it.
SICs are important practically, with applications to quantum
tomography~\cite{Ballester,Scott06,Petz,Zhu11}, quantum
cryptography~\cite{Englert,Rehacek,Renes05,Durt06,Durt08}, quantum
communication~\cite{Fuchs03,Fuchs04,Kim,Bodmann07,Oreshkov},
Kochen-Specker arguments~\cite{Bengtsson11b}, high precision
radar~\cite{Howard,Bodmann08,Hermann} and speech
recognition~\cite{Balan} (the last two applications being classical).
They are also important from a mathematical point of view, as giving
insight into the geometrical structure of quantum state
space~\cite{BengtssonBook}, and from a foundational point of view, as
playing a central role in the qbist approach to the interpretation of
quantum mechanics~\cite{Fuchs09,Fuchs10,Fuchs11}.  SICs have been
realized experimentally~\cite{Durt06,Medendorp}.  They have been
calculated numerically for every dimension $\le 67$ and exact
solutions have been constructed for dimensions $2$--$16$, $19$, $24$,
$35$ and $48$ (see Ref.~\cite{Scott10} for a comprehensive listing of
solutions known in 2010 and Ref.~\cite{monomial} for the exact
solution in dimension $16$).  There are therefore grounds for
conjecturing that SICs actually exist in every finite dimension.
However, in spite of strenuous efforts by many investigators over a
period of more than 10 years the question is still undecided.
The connection with the Clifford group comes from the fact that the overwhelming majority of known SICs are covariant with respect to the Weyl-Heisenberg group (or generalized Pauli group).  Moreover, Weyl-Heisenberg covariant SICs have been found in every dimension for which SICs have been found at all.  It is therefore tempting to conjecture that, not just SICs, but specifically Weyl-Heisenberg covariant SICs exist in every finite dimension.

The investigation reported here was motivated by two striking facts
concerning Weyl-Heisenberg covariant SICs.  Firstly, the defining
equations are massively over-determined.  Specifically, one has $d^2$
equations for only $2d-1$ real parameters. So if one did not know
better one might think it unlikely that solutions existed at all, in
any dimension $>2$.  The fact that solutions actually do exist (at
least numerically) in every dimension up to $67$, suggests the
presence of some kind of concealed symmetry or other algebraic feature
of the equations which forces a solution notwithstanding the
over-determination; and it further suggests that identifying that
feature might be the key to the existence problem.  The second
striking fact is that the known exact solutions all turn out to be
expressible in terms of radicals.  Again, this could not be
anticipated in advance of actually calculating the solutions, and
seeing that they are so expressible.  The defining equations of a
Weyl-Heisenberg covariant SIC fiducial vector are quartic polynomials
in the real and imaginary parts of the components.  The standard way
to solve such a system of equations (except in the simplest cases) is
to construct a Gr\"{o}bner basis~\cite{Cox}, thereby reducing the
problem to that of solving polynomial equations in a single variable.
It turns out~\cite{Grassl04,Grassl05,Grassl08,Scott10} that the single
variable equations are all of degree $\gg 4$ and, as is well-known,
such equations are only solvable in terms of radicals in the
exceptional case when the Galois group is solvable~\cite{Roman}.  In
fact, one finds\footnote{As originally noticed by Jon
  Yard~\cite{Yard}, although Ref.~\cite{Scott10} gives correct
  expressions for the exact fiducials in dimension $14$, there is an
  error in the calculation of the Galois group. The correct Galois
  group for this dimension can be found in
  Ref.~\cite{Appleby12}.}~\cite{monomial,Scott10,Appleby12} that in
all the known cases the Galois group has a particularly simple form
even amongst the class of solvable groups, having a subnormal series
of the form $\{e\}\vartriangleleft H \vartriangleleft G $ where $e$ is
the identity, $H$ and $G/H$ are Abelian, and $G/H$ is order $2$ ($G$
being the Galois group of the smallest normal extension of
$\mathbb{Q}$ containing all the components of the fiducial together
with $\sqrt{-1}$). It seems probable that this feature of the Galois
group is giving us an important clue, and we would naturally like to
investigate it further.  The trouble is that, although the Galois
group has a very simple structure, the solutions themselves are not
(in many cases the print out extends to many pages), which tends to
obstruct progress.  We were therefore led to wonder if a change of
basis would have the effect of simplifying the solutions.  In
Ref.~\cite{monomial} it was shown that the monomial representation
does indeed produce considerable simplification for $N=4$, $9$, $16$.
In the following we show that the $k$-nomial representation is
similarly efficacious.  We perform a hand calculation of the
Weyl-Heisenberg fiducials for $N=8$ and we find that the expressions
which result are indeed much simpler than the standard basis
expressions obtained by machine calculation~\cite{Scott10}.  The fact
that a hand calculation is even possible may be taken as further
evidence that the $k$-nomial basis is better adapted to the
problem. We also describe a machine calculation of a fiducial in
dimension $28$.  Although exact fiducials have previously been
calculated in dimensions $24$, $35$ and $48$, these relied on the
existence of additional symmetries of order $6$, $12$ and $24$,
respectively. So far, $N=28$ is the largest dimension in which it has
been possible to find an exact solution only with the help of a
symmetry of order $6$, while $N=16$ in Ref.~\cite{monomial} is the
largest dimension for which the computation relied only on a symmetry
of order three, as conjectured by Zauner. This may be taken to confirm
the hypothesis that the $k$-nomial basis is indeed better adapted to
the SIC problem.  As further confirmation of this point we revisited
the calculation of an exact fiducial in dimension~$12$ first reported
in Ref.~\cite{Grassl08}.  We found that the use of a $k$-nomial basis
reduced the computation time by more than $3$ orders of magnitude.  We
also found that the expressions for the fiducial are much more compact
than the ones given in Refs.~\cite{Grassl08,Scott10}

Finally, let us observe that the advantage of a hand calculation, such as the
solution for $N=8$ presented here, or the solution for $N=9$ presented in
Ref.~\cite{monomial}, is that it gives us a degree of insight into the algebraic
intricacies of the problem which a machine calculation cannot provide.  We
suggested above that the fact that solutions exist notwithstanding the massive
degree of over-determination, and the striking simplicity of the Galois group,
both hint at the presence of some underlying symmetry, or other algebraic
feature of the equations which has so far eluded us.  It may be that a study of
hand-constructed solutions will lead us to the secret which will enable us
finally to crack the problem.

The plan of the paper is as follows.  In Section~\ref{sec:rep} we describe the
$k$-nomial representation of the Clifford group.  In  Section~\ref{sec:sics} we briefly review a few basic facts about SICs.  In Section~\ref{sec:dim8}
give a hand calculation of the Weyl-Heisenberg SIC fiducials in dimension $8$ .  In Section~\ref{sec:dim12}
we obtain, by machine calculation,  a very compact solution for dimension $12$.  In Section~\ref{sec:dim28} we obtain, by machine
calculation, exact fiducials in dimension $28$.

\section{The $k$-nomial Representations}
\label{sec:rep}
To fix notations we begin by describing the standard basis
representation of the Clifford group.  For more details see (for
example) Ref.~\cite{Appleby05}.  Let $|0\rangle, \dots |N-1\rangle$ be
the standard basis in dimension $N$.  Define the operators $X$ and $Z$
by
\begin{align}
X |u\rangle & = |u+1\rangle \\ Z |u\rangle & = \omega^u |u\rangle
\end{align}
where addition of ket-labels is mod $N$ and $\omega= e^{\frac{2 \pi
    i}{N}}$.  The Weyl-Heisenberg displacement operators are then
defined by
\begin{equation}
D_{\mathbf{p}} = \tau^{p_1 p_2} X^{p_1} Z^{p_2}
\end{equation}
where $\mathbf{p} = \left(\begin{smallmatrix} p_1
  \\ p_2 \end{smallmatrix}\right)$, $\tau= -e^{\frac{\pi i}{N}}$ and
$p_1$, $p_2$ run from $0$ to $\bar{N}-1$, where
\begin{equation}
\bar{N} = \begin{cases} N \qquad & \text{$N$ odd} \\ 2 N \qquad &
  \text{$N$ even} \end{cases}
\end{equation}
With this definition
\begin{equation}
D_{\mathbf{p}}D_{\mathbf{q}} = \tau^{\langle
  \mathbf{p},\mathbf{q}\rangle} D_{\mathbf{p}+\mathbf{q}}
\label{eq:Dprd}
\end{equation}
where the symplectic form $\langle \cdot , \cdot \rangle$ is defined
by
\begin{equation}
\langle \mathbf{p}, \mathbf{q}\rangle = p_2q_1 -p_1 q_2
\end{equation}
The symplectic group $\SL(2,\mathbb{Z}_{\bar{N}})$ consists of all
matrices
\begin{equation}
F = \bmt \alpha & \beta \\ \gamma & \delta \emt
\end{equation}
such that $\alpha$, $\beta$, $\gamma$, $\delta \in
\mathbb{Z}_{\bar{N}}$ and $\Det F= 1$ (mod $\bar{N}$).  To each such
matrix there corresponds a unitary $U_F$, unique up to a phase, such
that
\begin{equation}
U\vpu{\dagger}_F D\vpu{\dagger}_{\mathbf{p}} U^{\dagger}_F =
D\vpu{\dagger}_{F\mathbf{p}}
\end{equation}
If $\beta$ is relatively prime to $\bar{N}$ we have the explicit
formula
\begin{equation}
U_F = \frac{e^{i\theta}}{\sqrt{N}} \sum_{u,v=0}^{N-1}
\tau^{\beta^{-1}\left(\delta u^2 -2 u v+\alpha v^2\right)} |u\rangle
\langle v |
\label{eq:UFdef}
\end{equation}
where $\beta^{-1}$ is the multiplicative inverse of $\beta$ (mod
$\bar{N}$) and $e^{i\theta}$ is an arbitrary phase.  Observe that it
follows from this formula that $U_F$ is a Hadamard matrix in the
standard basis whenever $\beta$ is relatively prime to $\bar{N}$.  If
$\beta$ is not relatively prime to $\bar{N}$ we use the
fact~\cite{Appleby05} that $F$ has the decomposition
\begin{equation}
F=F_1F_2 = \bmt \alpha_1 & \beta_1 \\ \gamma_1 & \delta_1 \emt \bmt
\alpha_2 & \beta_2 \\ \gamma_2 & \delta_2 \emt
\label{eq:primeDecomp}
\end{equation}
where $\beta_1$, $\beta_2$ both are relatively prime to $\bar{N}$ so
that $U_{F_1}$, $U_{F_2}$ can be calculated using
Eq.~(\ref{eq:UFdef}).  $U_F$ is then given by
\begin{equation}
U_F = U_{F_1} U_{F_2}
\end{equation}
up to an arbitrary phase.  The Clifford group then consists of all
products $D_{\mathbf{p}} U_F$.

The extended Clifford group is also important.  Enlarge
$\SL(2,\mathbb{Z}_{\bar{N}})$ to the group
$\ESL(2,\mathbb{Z}_{\bar{N}})$ by including the anti-symplectic
matrices
\begin{equation}
F = \bmt \alpha & \beta \\ \gamma & \delta \emt
\end{equation}
with $\Det F= -1$ (mod $\bar{N}$). Each such matrix has the unique
decomposition
\begin{equation}
F = \tilde{F} J
\end{equation}
where $\tilde{F}$ is symplectic and
\begin{equation}
J = \bmt 1 & 0 \\ 0 & -1\emt.
\label{eq:Jdef}
\end{equation}
To $J$ we associate the anti-unitary $U_J$ which acts by complex
conjugation in the standard basis:
\begin{equation}
U_J \left( \sum_{u=0}^{N-1} \psi_u |u\rangle \right) =
\sum_{u=0}^{N-1} \psi^{*}_u | u\rangle
\end{equation}
for all $|\psi\rangle = \sum_{u=0}^{N-1} \psi_u |u\rangle$.  We then
associate to $F$ the anti-unitary
\begin{equation}
U_{\vphantom{\tilde{F}}F} = U_{\tilde{F}} U_{\vphantom{\tilde{F}}J}
\label{eq:antiUDecomp}
\end{equation}
The extended Clifford group consists of all products of the form
$D_{\mathbf{p}} U_F$, with $\mathbf{p}\in \mathbb{Z}^2_{\bar{N}}$ and
$F \in \ESL(2,\mathbb{Z}_N)$.

These preliminaries completed we now turn to the construction of the
$k$-nomial representations.  Let $N = k n^2$ (one would usually choose
$k$ to be square-free, so as to make the non-zero blocks in the
representation matrices as small as possible; however this is not
necessary). Then it can be seen from Eq.~(\ref{eq:Dprd}) that $X^{kn}$
and $Z^{kn}$ commute, and are therefore simultaneously diagonalizable.
It is easily verified that a joint eigenbasis is
\begin{align}
|r,s,j\rangle = \frac{1}{\sqrt{n}} \sum_{t=0}^{n-1} \lambda^{-rt} |(s+jn)+tkn\rangle
\end{align} 
where $\lambda = e^{\frac{2\pi i}{n}}$, where $r,s\in\mathbb{Z}_n$
label the eigenspaces according to
\begin{align}
X^{kn} |r,s,j\rangle & = \lambda^r |r,s,j\rangle & Z^{kn} |r,s,j\rangle &  = \
\lambda^s |r,s,j\rangle
\end{align}
and where $j\in \mathbb{Z}_k$ labels the basis vectors within each eigenspace. 
The action of $X$, $Z$ in this basis is
\begin{align}
X|r,s,j\rangle & = 
\begin{cases}
|r,s+1,j\rangle \qquad & s\neq n-1
\\
|r,s+1,j+1\rangle \qquad & s= n-1,\ j\neq k-1
\\
\lambda^r | r,s+1,j+1\rangle \qquad & s=n-1,\ j= k-1
\end{cases}
\\
Z|r,s,j\rangle & = \omega^{s+n j} |r-1,s,j\rangle\label{eq:Z_rsj}
\end{align}
from which we see  that the displacement operators are $k$-nomial in
this basis (in fact monomial) just as they are in the standard basis.
From Eq.~(\ref{eq:Z_rsj}) it follows that not only $X^{kn}$ and
$Z^{kn}$ are diagonal in this basis, but also $Z^n$.

The advantage of this basis is that in it, not only the displacement
operators, but also the symplectic unitaries are represented by
$k$-nomial matrices.  In fact, let
\begin{equation}
F = \bmt \alpha & \beta \\ \gamma & \delta \emt
\end{equation}
be an arbitrary matrix $\in \SL(2,\mathbb{Z}_{\bar{N}})$.  Then
\begin{align}
U^{\dagger}_F X^{kn} U\vpu{\dagger}_F 
& =
\begin{cases}
(-1)^{\gamma \delta} X^{kn\delta} Z^{-k n\gamma} \qquad & \text{$k$ odd, $n$ even}
\\
X^{kn\delta} Z^{-kn\gamma} \qquad & \text{otherwise}
\end{cases}
\\
U^{\dagger}_F Z^{kn} U\vpu{\dagger}_F 
& =
\begin{cases}
(-1)^{\alpha \beta} X^{-kn\beta} Z^{k n\alpha} \qquad & \text{$k$ odd, $n$ even}
\\
X^{-kn\beta} Z^{k n\alpha} \qquad & \text{otherwise}
\end{cases}
\end{align}
 where we used the fact that
\begin{equation}
\tau^{k^2 n^2} = \begin{cases} 1 \qquad & \text{$N$ odd} \\ (-1)^k \qquad & \text{$N$ even} \end{cases}
\end{equation}
to calculate the signs.  Consequently
\begin{align}
X^{kn} U_F |r,s,j\rangle & =
\lambda^{r'}U_F |r,s,j\rangle 
\\
Z^{kn} U_F |r,s,j\rangle & =
\lambda^{s'}U_F |r,s,j\rangle 
\end{align}
where
\begin{align}
r'& = \delta r -\gamma s + m \gamma \delta
\label{eq:rPrimeDef}
\\
s' & = -\beta r +\alpha s + m \alpha \beta
\label{eq:sPrimeDef}
\\
\intertext{with}
m & = \begin{cases} \frac{n}{2} \qquad & \text{$k$ odd, $n$ even,}
\\
0 \qquad & \text{otherwise.}
\end{cases}
\end{align}
So $U_F$ takes the $(r,s)$ eigenspace to the $(r',s')$ eigenspace:
\begin{equation}
U_F |r,s,j\rangle = \sum_{j'=0}^{k-1} \left(M_{F,rs}\right)_{jj'}|r',s',j'\rangle
\end{equation}
for some family of $k\times k$ matrices $M_{F,rs}$.  It is thus
$k$-nomial, as claimed. Note that this also shows that the eigenspaces
labeled by $(r,s)$ form a system of imprimitivity (see
Eq.~(\ref{eq:ImprimSystem})).

To calculate the matrices $M_{F,rs}$ suppose, first, that $\beta$ is
relatively prime to $\bar{N}$.  Then using Eq.~(\ref{eq:UFdef}) one
finds, after a certain amount of algebra, that, up to an overall
phase,
\begin{equation}
\left(M_{F,rs}\right)_{jj'}
=
\frac{1}{\sqrt{k}}\tau^{\beta^{-1}\left(\delta (s'+j'n)^2 -2(s'+j'n)(s+j n) + \alpha(s+ jn)^2\right)}
\end{equation}
where $s'$ is given in terms of $r$, $s$ by Eq.~(\ref{eq:sPrimeDef})
(note that in this formula it is essential that $s'$ be reduced mod
$n$, so that it is an integer in the range $0\le s' < n$).  The case
when $\beta$ is not relatively prime to $\bar{N}$ can then be handled
using the decomposition of Eq.~(\ref{eq:primeDecomp}).

It is perhaps interesting to note that when $\beta$ is relatively prime to
$\bar{N}$ the non-zero blocks in the $k$-nomial representation are Hadamard (in
contrast to the standard representation where the whole matrix is Hadamard).

Finally, we can calculate the anti-symplectic anti-unitaries using
Eq.~(\ref{eq:antiUDecomp}) together with the fact
\begin{equation}
U_J |r,s,j\rangle = |-r,s,j\rangle
\end{equation}

\section{SICs}
\label{sec:sics}
The $k$-nomial representations described in the last section are a
very general construction which is potentially relevant to any area
where the Clifford and extended Clifford groups play a role.  In the
remainder of this paper we illustrate their use by showing how they
can be used to simplify the calculation of SICs.  We begin, in this
section, by reviewing a few basic facts (for more details see
Refs.~\cite{Appleby05,Renes04,Zauner,Scott10} ).

A SIC in dimension $N$ is a family of $N^2$ operators
\begin{equation}
E_r = \frac{1}{N}|\psi_r\rangle\langle \psi_r |
\end{equation}
such that the vectors $|\psi_r\rangle$ satisfy
\begin{equation}
\bigl| \langle \psi_r | \psi_s \rangle \bigr|= \begin{cases}1 
\qquad & r=s,
\\
\frac{1}{\sqrt{N+1}} \qquad & \text{otherwise.}
\end{cases}
\end{equation}
The fact that the matrix $\Tr(E_rE_s)$ is non-singular means that the
$E_r$ are a basis for operator space and
\begin{equation}
\sum_{r=1}^{N^2} E_r = I.
\end{equation}
So the $E_r$ form an informationally complete POVM (positive operator valued measure).  

The vast majority of SICs which have been constructed to date are covariant under the action of the Weyl-Heisenberg group.  That is, they have the property
\begin{equation}
D^{\vphantom{\dagger}}_{\mathbf{q}} E_{\mathbf{p}} D^{\dagger}_{\mathbf{q}} = E_{\mathbf{p}+\mathbf{q}}
\end{equation}
for all $\mathbf{p}$, $\mathbf{q}$ (where, instead of labelling  the POVM elements by the single index $r$, we label them by $\mathbf{p}\in\mathbb{Z}_N^2$).  For a SIC of this  kind we can generate the entire SIC by acting on $E_{\boldsymbol{0}}$ with the displacement operators $D_{\mathbf{p}}$.  The vector $|\psi_{\boldsymbol{0}}\rangle$ corresponding to $E_{\boldsymbol{0}}$ is called the fiducial vector.  We usually omit the subscript and simply denote the fiducial vector $|\psi\rangle$.  The problem of constructing a Weyl-Heisenberg covariant SIC thus reduces to the problem of finding a single vector $|\psi\rangle$ such that 
\begin{equation}
\bigl| \langle \psi | D_{\mathbf{p}} |\psi\rangle\bigr| =
\begin{cases}
1 \qquad & \mathbf{p} = \boldsymbol{0} \mod N
\\
\frac{1}{\sqrt{N+1}} \qquad &  \mathbf{p} \neq \boldsymbol{0} \mod N
\end{cases}
\end{equation}

It is a striking,  so far unexplained fact that every known Weyl-Heisenberg SIC fiducial vector is an eigenvector of a Clifford unitary  $D_{\mathbf{q}} U_F$ for which $\Tr F = -1 \mod N$ and $F\neq I$ (note that the requirement $F\neq I$ is automatic if $N\neq 3$).  It can be shown~\cite{Zauner,Appleby05,Scott10} that such unitaries are always order 3 up to a phase.  Conversely, it appears that  every such  unitary has a Weyl-Heisenberg SIC fiducial as one of its eigenvectors.    In the following sections we exploit this fact by looking for fiducials which are eigenvectors of $U_{F_z}$, where
\begin{equation}
F_Z = \bmt 0 & -1 \\ 1 & -1\emt
\end{equation}
is the Zauner matrix~\cite{Zauner}.  We refer to $U_{F_z}$ as the Zauner unitary. 
\section{Dimension $8$ fiducial}
\label{sec:dim8}

We now perform a hand-calculation of the fiducials in dimension $8$
which are covariant under the version of the Weyl-Heisenberg group
considered in this paper (for fiducials covariant with respect to a
different version of the group see Refs.~\cite{Hoggar,Godsil}).  Of
course exact expressions for the standard basis components have
already been found by Scott and Grassl~\cite{Scott10} (by means of a
machine-calculation of a Gr\"{o}bner basis).  So if we only wanted to
make the point that the expressions become much simpler in the
$k$-nomial basis it would be enough just to transform the expressions
in Ref.~\cite{Scott10}.  However, as we stated in the Introduction, we
feel that a hand-calculation gives a greater degree of insight into
the problem than one gets from a machine-calculation. In particular we
entertain the hope that a close inspection of hand-calculated
solutions may enable us to spot the ``secret ingredient'' responsible
for the fact that SICs exist (at least numerically for dimensions $\le
67$) in spite of the massive degree of overdetermination, and for the
striking features of the Galois group---thereby, perhaps, leading to a
solution of the existence problem.

It may be worth remarking that we found it surprising that we were
able to solve the equations at all (by hand, that is).  One of us
(DMA) well recalls that he could only obtain the dimension 7 fiducials
in Ref.~\cite{Appleby05} by a process of trial and error, and that in
order to be sure that he had found all the solutions he had to appeal
to the numerical work of Renes \emph{et al.}~\cite{Renes04}.

We look for SIC fiducial vectors which are eigenvectors of the Zauner
unitary, $U_{F_z}$.  Note that $8$ is one of the special
dimensions~\cite{Scott10} for which fiducials can be found in all
three eigenspaces of $U_{F_z}$.  The first step in the calculation is
to choose a basis which diagonalizes $U_{F_Z}$.  Since there are only
$3$ eigenspaces there are many ways to do this, which are far from
being equivalent from the point of view of calculational complexity.
One natural way to proceed, which causes the equations to be solved to
take a particularly simple form, is to consider the normalizer
$\mathcal{N}_Z$ of the cyclic subgroup generated by
$F_Z$---\emph{i.e.}\ the set of matrices $G\in
\ESL(2,\mathbb{Z}_{\bar{N}})$ such that
\begin{equation}
G F_Z = F^r_Z G
\end{equation}
for $r=1$ or $2$.  The significance of $\mathcal{N}_Z$ is that the
corresponding unitaries and anti-unitaries permute the eigenvectors of
$U_{F_Z}$.  In particular they permute the fiducials which are
eigenvectors of $U_{F_Z}$.  Having constructed $\mathcal{N}_Z$ one
then picks out a maximal subset containing $F_Z$ such that the
corresponding unitaries/anti-unitaries commute.  The joint
eigenvectors of the operators in this subgroup give us our desired
basis.  As we will see the fact that the matrices are all $k$-nomial
considerably facilitates the calculation.

The group $\mathcal{N}_Z$ is generated by the matrices
\begin{align}
K & = \bmt 0 & 1 \\ 1 & 0 \emt & 
A & = \bmt 1 & 5\\ -5 & 6\emt& 
P & = \bmt -1 & 0 \\ 0 & -1 \emt
\end{align}
Here $K$ is order $2$ anti-symplectic, $A$ is order $24$ anti-symplectic
and $P$ is order $2$ symplectic.  We have
\begin{equation}
F_Z = A^8
\end{equation}
and
\begin{align}
PA & = A P & P K & = K P & A K & = K A^{-1} P
\end{align}
Define symplectic matrices
\begin{align}
\tilde{K} & = \bmt 0 & -1 \\ 1 & 0 \emt &
\tilde{A} & = \bmt 1 & -5 \\ -5 & - 6\emt
\end{align} 
We then have 
\begin{align}
K & = \tilde{K} J & A & = \tilde{A} J
\end{align}
and, consequently,
\begin{align}
U_K & = U_{\tilde{K}} U_{\vphantom{\tilde{K}}J} & U_A & = U_{\tilde{A}} U_{\vphantom{\tilde{K}}J}
\end{align}
where $J$ is the matrix defined in Eq.~(\ref{eq:Jdef}).

We now calculate the matrices corresponding to these operators.  For
dimension $8$ we have $k=n=2$.  If we order the basis vectors
$|r,s,j\rangle$ lexicographically
\begin{equation}
|0,0,0\rangle , |0,0,1\rangle, |0,1,0\rangle, \dots ,|1,1,1\rangle
\end{equation}
the matrices become (with a given choice for the arbitrary phase factors) 
\begin{align}
U_{\tilde{K}} & =
\bmt
K_0 & 0 & 0 & 0 \\ 0 & 0 &  K_1 & 0 \\ 0 & K_2 & 0 & 0 \\ 0 & 0 & 0 & K_3 
\emt
&
U_{\tilde{A}} & = 
\bmt
A_0 & 0 & 0 & 0 \\ 0 & 0 & A_1 & 0 \\ 0 & 0 & 0 & A_2 \\ 0 & A_3 & 0 & 0 
\emt
\end{align}
and
\begin{equation}
U_P  = 
\bmt
I & 0 & 0 & 0 \\ 0 & \sigma_x &  0 & 0 \\ 0 & 0 & \sigma_z & 0 \\ 0 & 0 & 0 & -\sigma_x
\emt
\end{equation}
where $\sigma_x$, $\sigma_z$ are the Pauli matrices and
\begin{align}
K_0 & = \frac{1}{\sqrt{2}}
\bmt 1 & 1 \\ 1 & -1 \emt
&
A_0 & = \frac{1}{\sqrt{2}}
\bmt 1 & -i \\ -1 & -i 
\emt
\\
K_1 & = \frac{1}{\sqrt{2}} 
\bmt 1 & i\\ 1 & -i 
\emt
&
A_1 & = \frac{
\tau^{-2}}{\sqrt{2}}
\bmt 1  & 1 \\ 1 & -1
\emt
\\
K_2 & = \frac{1}{\sqrt{2}}
\bmt 1 & 1 \\ i & -i
\emt
&
A_2 & = \frac{\tau^{3}}{\sqrt{2}}
\bmt 1 & -1 \\ -i &  -i
\emt
\\
K_3 & = \frac{\tau^2}{\sqrt{2}}
\bmt 1 & i \\ i & 1
\emt
&
A_3 & = \frac{\tau^{-5}}{\sqrt{2}}
\bmt
1 & -i \\ i & -1
\emt
\end{align}
We fix the arbitrary phase in the definition of $U_{F_Z}$ by making the choice
\begin{equation}
U_{F_Z} = U^8_A
\end{equation}
It is readily confirmed that $U_K$ and $U_P$ are order 2.  However
$U_A$, unlike $A$, is only order $12$.  This is
because~\cite{Appleby05} the mapping $G \mapsto U_G$ takes $A^{12} =
\left(\begin{smallmatrix} 9 & 0 \\ 0 & 9 \end{smallmatrix}\right)$ to
a multiple of the identity.

We seek the joint eigenvectors of the commuting unitaries $U\vpu{2}_P$
and $U_A^2$.  We have
\begin{equation}
U_A^2 = U\vpu{*}_{\tilde{A}}U^{*}_{\tilde{A}} = 
\bmt 
A_{00} & 0 & 0 & 0 \\ 
0  & 0 & 0 & A_{12}  \\ 
0 & A_{23} & 0 & 0 \\
0 & 0 & A_{31} & 0 
\emt
\end{equation}
where
\begin{equation}
A_{rs} = A\vpu{*}_r A^{*}_s
\end{equation}
The $2\times 2$ block $A_{00}$ is easily diagonalized.  To diagonalize the other, $6\times 6$ diagonal block of $U^2_A$ observe that
\begin{equation}
A_{12}A_{31}A_{23} = A_{23}A_{12}A_{31} = A_{31}A_{23}A_{12} =I 
\end{equation}
Consequently
\begin{equation}
\bmt 0 & 0 & A_{12} \\ A_{23} & 0 & 0 \\ 0 & A_{31} & 0 \emt 
\bmt u_1 \\ u_2 \\ u_3 \emt = \xi \bmt u_1 \\ u_2 \\ u_3 \emt
\end{equation}
with $\xi \ne 0$ if and only if $\xi$ is a cube root of unity and
\begin{align}
u_1 & = \xi A_{12}A_{31} u_2 & u_3 & = \xi^2 A_{31} u_2
\end{align}
If we impose the further requirement that
\begin{equation}
\bmt 0 \\ u_1 \\ u_2 \\ u_3\emt
\end{equation}
be an eigenvector of $U_P$ we must have that $u_2$ is a multiple of
$\left(\begin{smallmatrix} 1 \\ 0 \end{smallmatrix}\right)$ or
$\left(\begin{smallmatrix} 0 \\ 1 \end{smallmatrix}\right)$.  We
conclude that a complete set of joint eigenvectors is 
\begin{align}
|0,1\rangle & = 
\frac{1}{\sqrt{6}}
\begin{pmatrix}
0, & 0, & \eta^{33}, & \eta^{33}, & \sqrt{2} \eta^{15}, & 0, & -1, & 1 
\end{pmatrix}^t
\\
|0,-1\rangle & = 
\frac{1}{\sqrt{6}}
\begin{pmatrix}
0, & 0, & \eta^{27}, & \eta^{3}, & 0, & \sqrt{2} \eta^{33}, & \eta^{6}, & \eta^{6}
\end{pmatrix}^t
\\
|1,1\rangle & = 
\frac{1}{6}
\begin{pmatrix}
\eta^{23} \sqrt{6(3+\sqrt{3})}, & \eta^{29} \sqrt{6(3-\sqrt{3})}, & 0, & 0, & 0, & 0, & 0, & 0
\end{pmatrix}^t
\\
|2,1\rangle & = 
\frac{1}{\sqrt{6}}
\begin{pmatrix}
0, & 0, & \eta^{17}, & \eta^{17}, & \sqrt{2} \eta^{31}, & 0 , & -1, & 1
\end{pmatrix}^t
\\
|2,-1\rangle & = 
\frac{1}{\sqrt{6}}
\begin{pmatrix}
0, & 0, & \eta^{11}, & \eta^{35}, & 0, & \sqrt{2} \eta, & \eta^{6}, & \eta^{6}
\end{pmatrix}^t
\\
|4,1\rangle & = 
\frac{1}{\sqrt{6}}
\begin{pmatrix}
0, & 0, & \eta, & \eta, & \sqrt{2} \eta^{47}, & 0 , & -1, & 1
\end{pmatrix}^t
\\
|4,-1\rangle & = 
\frac{1}{\sqrt{6}}
\begin{pmatrix}
0, & 0, & \eta^{43}, & \eta^{19}, & 0, & \sqrt{2} \eta^{17}, & \eta^{6}, & \eta^{6}
\end{pmatrix}^t
\\
|5,1\rangle & = \frac{1}{6}
\begin{pmatrix}
\eta^{7} \sqrt{6(3-\sqrt{3})}, & \eta^{37} \sqrt{6(3+\sqrt{3})}, & 0, & 0, & 0, & 0, & 0, & 0
\end{pmatrix}^t & &
\end{align}
where $\eta= e^{\frac{\pi i}{24}}$,
\begin{align}
U^2_A |r,s\rangle & = \eta^{8r}|r,s\rangle & U_P |r,s\rangle & = s |r,s\rangle
\end{align}
for all $r,s$, and where the phases have been chosen so that
\begin{equation}
U_K |r,s\rangle = |r,s\rangle
\label{eq:UKEvecCond}
\end{equation}
for all $r,s$.  Note that the fact that $U_K$ is an anti-unitary means
that Eq.~(\ref{eq:UKEvecCond}) fixes the phase of $|r,s\rangle$ up to
a sign (this is an important point:  if one chooses the phases
differently the fiduciality conditions are much harder to solve). The
action of the anti-unitary $U_A$ is: 
\begin{align}
U_A |0,1\rangle & = \eta^{42} |0,1\rangle  & U_A|0,-1\rangle & = \eta^{30}|0,-1\rangle
\\
U_A |1,1\rangle & = \eta^{14}|5,1\rangle & &
\\
U_A |2,1\rangle & =  \eta^{10}|4,1 \rangle & U_A |2,-1\rangle & = \eta^{46} |4,-1\rangle
\\
U_A | 4,1\rangle & = \eta^{26} |2,1\rangle & U_A | 4,-1\rangle & = \eta^{14} |2,-1\rangle
\\
U_A |5,1\rangle & = \eta^{22} |1,1\rangle
\end{align} 
The eigenspaces of $U\vpu{8}_{F_Z} = U_A^8$ are as in Table~\ref{tble:espaces}.
\begin{table}[hbt]
\centerline{\footnotesize
\begin{tabular}{|c|c|c|c|}
\hline $\myStrut{4}$
\te{1.8}{eigenspace} & \te{1.8}{eigenvalue} & \te{1.8}{dimension} & \te{2.5}{basis}   \\
\hline
$\mathcal{S}_0$ &   $\myStrut{4}1$ & 2 & $|0,1\rangle$, $|0,-1\rangle$ \\
\hline
$\mathcal{S}_1$ &   $\myStrut{4} \eta^{16}$ & 3 & $|5,1\rangle$, $|2,1\rangle$, $|2,-1\rangle$\\
\hline
$\mathcal{S}_2$ &   $\myStrut{4} \eta^{32}$ & 3 & $|1,1\rangle$, $|4,1\rangle$, $|4,-1\rangle$\\
\hline
\end{tabular}}
\caption{\label{tble:espaces}}
\end{table}
As we remarked above $N=8$ is one of the dimensions for which
fiducials exist in all $3$ eigenspaces of the Zauner unitary.  Since
$U_A$ toggles the eigenspaces $\mathcal{S}_1$ and $\mathcal{S}_2$ we
only need to calculate fiducials in $\mathcal{S}_0$ and $\mathcal{S}_1$.
We thus need to solve the equations
\begin{equation}
\left| \langle \psi |D_{\mathbf{p}} | \psi \rangle \right|^2 = \frac{1}{9}
\end{equation}
for all $\mathbf{p} \ne 0$ (mod $8$) and $|\psi\rangle$ of the form
\begin{align}
|\psi\rangle & = \cos\theta |5,1\rangle + \sin \theta \cos\phi e^{i\chi_2} |2,1\rangle + \sin \theta \sin\phi e^{i \chi_3} |2,-1\rangle
\label{eq:fidS1}
\\
\intertext{with $0 \le \theta, \phi  \le \frac{\pi}{2}$ and $0 \le \chi_2, \chi_3 < 2 \pi$, or}
|\psi\rangle & =  \cos\theta |0,1\rangle + \sin \theta e^{i\chi} |0,-1\rangle
\label{eq:fidS0}
\end{align}
with $0 \le \theta \le \frac{\pi}{2}$ and $0 \le \chi < 2 \pi$.  The
fact that $|\psi\rangle$ is assumed to be an eigenvector of $U_{F_Z}$
means that it is enough to require that it satisfy the 11 equations
\begin{align}
\bigl|\langle \psi |D_{\mathbf{p}_1}\psi\rangle\bigr|^2 &= \frac{1}{9}
\\
\bigl|\langle \psi |D_{A^r\mathbf{p}_2}\psi\rangle\bigr|^2 &= \frac{1}{9} & r & = 0,1
\\
\bigl|\langle \psi |D_{A^r\mathbf{p}_3}\psi\rangle\bigr|^2 &= \frac{1}{9} & r & = 0,\dots,3
\\
\bigl|\langle \psi |D_{A^r\mathbf{p}_4}\psi\rangle\bigr|^2 &= \frac{1}{9} & r & = 0,\dots,3
\end{align}
with
\begin{align}
\mathbf{p}_1 &= (0,4) & \mathbf{p}_2 & = (0,2) & \mathbf{p}_3 & = (0,1) & \mathbf{p}_4 & = (1,2)
\end{align}
or, equivalently, the following linear combinations of these equations:
\begin{align}
\bigl|\langle \psi |D_{\mathbf{p}_2}\psi\rangle\bigr|^2 + \bigl|\langle \psi |D_{A\mathbf{p}_2}\psi\rangle\bigr|^2 & = \frac{2}{9}
\tag{E1}
\label{eq:E1}
\\
2\bigl|\langle \psi |D_{\mathbf{p}_1}\psi\rangle\bigr|^2-\bigl|\langle \psi |D_{\mathbf{p}_2}\psi\rangle\bigr|^2 - \bigl|\langle \psi |D_{A\mathbf{p}_2}\psi\rangle\bigr|^2 & =0 
\tag{E2}
\label{eq:E2}
\\
\sum_{r=0}^{3} k^{(1)}_r \bigl|\langle \psi |D_{A^r\mathbf{p}_3}\psi\rangle\bigr|^2
+
\sum_{r=0}^{3} k^{(1)}_r \bigl|\langle \psi |D_{A^r\mathbf{p}_4}\psi\rangle\bigr|^2
& = \frac{8}{9}
\tag{E3}
\label{eq:E3}
\\
\bigl|\langle \psi |D_{\mathbf{p}_2}\psi\rangle\bigr|^2 - \bigl|\langle \psi |D_{A\mathbf{p}_2}\psi\rangle\bigr|^2 & = 0
\tag{E4}
\label{eq:E4}
\\
\sum_{r=0}^{3} k^{(2)}_r \bigl|\langle \psi |D_{A^r\mathbf{p}_3}\psi\rangle\bigr|^2
-
\sum_{r=0}^{3} k^{(2)}_r \bigl|\langle \psi |D_{A^r\mathbf{p}_4}\psi\rangle\bigr|^2
& = 0
\tag{E5}
\label{eq:E5}
\\
\sum_{r=0}^{3} (\sqrt{3}k^{(2)}_r -k^{(1)}_r) \bigl|\langle \psi |D_{A^r\mathbf{p}_3}\psi\rangle\bigr|^2
+
\sum_{r=0}^{3}  (\sqrt{3}k^{(2)}_r +k^{(1)}_r)\bigl|\langle \psi |D_{A^r\mathbf{p}_4}\psi\rangle\bigr|^2
& = 0
\tag{E6}
\label{eq:E6}
\\
\sum_{r=0}^{3} (k^{(2)}_r+\sqrt{3}k^{(1)}_r )\bigl|\langle \psi |D_{A^r\mathbf{p}_3}\psi\rangle\bigr|^2
+
\sum_{r=0}^{3} (k^{(2)}_r -\sqrt{3}k^{(1)}_r \bigl|\langle \psi |D_{A^r\mathbf{p}_4}\psi\rangle\bigr|^2
& = 0
\tag{E7}
\label{eq:E7}
\\
\sum_{r=0}^{3} k^{(3)}_r \bigl|\langle \psi |D_{A^r\mathbf{p}_3}\psi\rangle\bigr|^2
& = 0
\tag{E8}
\label{eq:E8}
\\
\sum_{r=0}^{3} k^{(3)}_r \bigl|\langle \psi |D_{A^r\mathbf{p}_4}\psi\rangle\bigr|^2
& = 0
\tag{E9}
\label{eq:E9}
\\
\sum_{r=0}^{3} k^{(4)}_r \bigl|\langle \psi |D_{A^r\mathbf{p}_3}\psi\rangle\bigr|^2
& = 0
\tag{E10}
\label{eq:E10}
\\
\sum_{r=0}^{3} k^{(4)}_r \bigl|\langle \psi |D_{A^r\mathbf{p}_4}\psi\rangle\bigr|^2
& = 0
\tag{E11}
\label{eq:E11}
\end{align} 
where
\begin{align}
k^{(1)} & = \bmt 1 \\ 1 \\ 1 \\ 1 \emt
&
k^{(2)} &= \bmt 1 \\ -1 \\ 1 \\ -1 \emt
&
k^{(3)} &= \bmt 1 \\ 0 \\ -1 \\ 0 \emt
&
k^{(4)} & = \bmt 0 \\ 1 \\ 0 \\ -1 \emt
\end{align}
We begin with the fiducials in $\mathcal{S}_1$.  Substituting
Eq.~(\ref{eq:fidS1}) into these equations we find that
Eqs.~(\ref{eq:E1})--(\ref{eq:E3}) are each equivalent to the single
condition
\begin{equation}
\cos2\theta(1+\cos2\theta) = 0
\end{equation}
The solution $\theta = \frac{\pi}{2}$ is inconsistent with the
remaining equations.  On the other hand if one sets $\theta=
\frac{\pi}{4}$ in Eqs.~(\ref{eq:E4})--(\ref{eq:E11}) it is
straightforward (though perhaps a little tedious) to show that the
resulting equations are equivalent to the four conditions
\begin{align}
\sin^2 2\phi \cos 2(\chi_2-\chi_3) & =- 2\sqrt{3}\cos 2\phi
\\
\cos^2\phi \cos 2\chi_2 &= \frac{1}{4\sqrt{6}}
\left(
\cos^2 2\phi +2\cos 2\phi -1
\right)
\\
\sin^2\phi \cos 2\chi_3 & = -\frac{1}{4\sqrt{2}}
\left(
\cos^2 2\phi -2\cos 2\phi -1
\right)
\\
\cos^2 2\phi -4\cos 2\phi -1
& = 0
\end{align}
whose solution is
\begin{align}
\cos \phi & = \sqrt{\frac{3-\sqrt{5}}{2}} 
\\
e^{i\chi_2} & = \frac{s_2}{4} \left(\sqrt{8+\sqrt{6}-\sqrt{30}}+ i s_1\sqrt{8-\sqrt{6}+\sqrt{30}} \right)
\\
e^{i\chi_3} & =\frac{s_3}{4} e^{-\frac{s_1\pi i}{12}} \left(\sqrt{8+\sqrt{6}-\sqrt{30}}+ i s_1\sqrt{8-\sqrt{6}+\sqrt{30}} \right)
\end{align}
where $s_1,s_2,s_3 = \pm 1$ can be chosen independently.

Turning to the fiducials in $\mathcal{S}_0$ we find on substituting
Eq.~(\ref{eq:fidS0}) into Eqs.~(\ref{eq:E1})--(\ref{eq:E11}) that the
equations reduce to the two conditions
\begin{align}
\sin^2 2\theta \cos 2\phi & = 0
\\
1+\cos 2\theta-\cos^2 2\theta  & = 0
\end{align}
whose solution is
\begin{align}
\cos \theta & = \frac{1}{2}\sqrt{3-\sqrt{5}}  & e^{i\phi} & =e^{\frac{(2r+1)\pi i}{4}}
\end{align}
where $r=0,\dots,3$. 

To conclude:   there are $8$ fiducials in the $\mathcal{S}_1$  eigenspace given by
\begin{equation}
|\psi\rangle = \frac{1}{2\sqrt{3}} 
\bmt 
\eta^7 \sqrt{3-\sqrt{3}} 
\\
\eta^{37} \sqrt{3+\sqrt{3}}
\\ 0 \\ 0 \\ 0 \\ 0 \\ 0 \\ 0
\emt
+\frac{s_2}{2}\sqrt{\frac{3-\sqrt{5}}{6}} e^{is_1\chi}
\bmt
0 \\ 0 \\ \eta^{17} \\ \eta^{17} \\ \sqrt{2}\eta^{31} \\ 0 \\ -1 \\ 1
\emt
+\frac{s_3}{2}\sqrt{\frac{\sqrt{5}-1}{6}} e^{is_1\left(\chi-\frac{\pi}{12}\right)}
\bmt
0 \\ 0 \\ \eta^{11} \\ \eta^{35} \\ 0 \\ \sqrt{2}\eta\\
\eta^{6} \\ \eta^{6}
\emt
\end{equation}
with $s_1,s_2,s_3=\pm 1$ and
\begin{equation}
e^{i\chi} =\frac{1}{4} \left(\sqrt{8+\sqrt{6}-\sqrt{30}}+ i \sqrt{8-\sqrt{6}+\sqrt{30}} \right)
\end{equation}
another $8$ fiducials in the $\mathcal{S}_2$ eigenspace obtained by
acting on these with the anti-unitary $U_A$, and $4$ fiducials in the
$\mathcal{S}_0$ eigenspace given by
\begin{equation}
|\psi\rangle =
\frac{1}{2}\sqrt{\frac{3-\sqrt{5}}{6}}
\bmt 0 \\ 0 \\ \eta^{33} \\ \eta^{33} \\ \sqrt{2}\eta^{15} \\ 0 \\ -1 \\ 1
\emt
+
\frac{i^r}{2}\sqrt{\frac{1+\sqrt{5}}{6}} 
\bmt 0 \\ 0 \\ \eta^{33} \\ \eta^9 \\ 0 \\ \sqrt{2} \eta^{39} \\ \eta^{12} \\ \eta^{12} 
\emt
\end{equation}
where $r=0, \dots ,3$.  The fiducials in $\mathcal{S}_1$,
$\mathcal{S}_2$ are on Scott-Grassl~\cite{Scott10} orbit $8a$; the
fiducials in $\mathcal{S}_0$ are on Scott-Grassl orbit $8b$.

Comparing these $k$-nomial basis expressions with the standard basis
expressions in Ref.~\cite{Scott10} we see that the degree of
simplification achieved is very considerable.

\section{Dimension $12$ fiducial}
\label{sec:dim12}
Before we present our new solution for dimension $28$, we revisit
dimenion $N=12=2^2\times 3$ to further demonstrate the advantage of
the sparse representation of the Clifford group. 

Choosing a basis such that both $X^6$ and $Z^6$ are diagonal, the
representation of the Clifford group will be $3$-nomial.  In this
basis, the Weyl-Heisenberg group is generated by
\def\w{\omega_{24}}
\begin{align}
X&=\left(
\begin{array}{*{12}{c}}
 0   &\w^{9} &  0   &  0   &  0   &  0   &  0   &  0   &  0   &  0   &  0   & 0\\
 0   &  0   &\w^{23}&  0   &  0   &  0   &  0   &  0   &  0   &  0   &  0   & 0\\
 0   &  0   &  0   &\w    &  0   &  0   &  0   &  0   &  0   &  0   &  0   & 0\\
 0   &  0   &  0   &  0   &\w^{15}&  0   &  0   &  0   &  0   &  0   &  0   & 0\\
 0   &  0   &  0   &  0   &  0   &\w^{17}&  0   &  0   &  0   &  0   &  0   & 0\\
\w^{7}&  0   &  0   &  0   &  0   &  0   &  0   &  0   &  0   &  0   &  0   & 0\\
 0   &  0   &  0   &  0   &  0   &  0   &  0   &\w^{9} &  0   &  0   &  0   & 0\\
 0   &  0   &  0   &  0   &  0   &  0   &  0   &  0   &\w^{11}&  0   &  0   & 0\\
 0   &  0   &  0   &  0   &  0   &  0   &  0   &  0   &  0   &\w^{17}&  0   & 0\\
 0   &  0   &  0   &  0   &  0   &  0   &  0   &  0   &  0   &  0   &\w^{11}& 0\\
 0   &  0   &  0   &  0   &  0   &  0   &  0   &  0   &  0   &  0   &  0   &\w\\
 0   &  0   &  0   &  0   &  0   &  0   &\w^{11}&  0   &  0   &  0   &  0   & 0
\end{array}
\right)\label{eq:X12}\\
\noalign{and}
Z&=\left(
\begin{array}{*{12}{c}}
 0   &  0   &  0   &  0   &  0   &  0   &\w^{15}&  0   &  0   &  0   &  0   & 0\\
 0   &  0   &  0   &  0   &  0   &  0   &  0   &\w^{17}&  0   &  0   &  0   & 0\\
 0   &  0   &  0   &  0   &  0   &  0   &  0   &  0   &\w^{7} &  0   &  0   & 0\\
 0   &  0   &  0   &  0   &  0   &  0   &  0   &  0   &  0   &\w    &  0   & 0\\
 0   &  0   &  0   &  0   &  0   &  0   &  0   &  0   &  0   &  0   &\w^{23}& 0\\
 0   &  0   &  0   &  0   &  0   &  0   &  0   &  0   &  0   &  0   &  0   &\w^{9}\\
\w^{9}&  0   &  0   &  0   &  0   &  0   &  0   &  0   &  0   &  0   &  0   & 0\\
 0   &\w^{11}&  0   &  0   &  0   &  0   &  0   &  0   &  0   &  0   &  0   & 0\\
 0   &  0   &\w    &  0   &  0   &  0   &  0   &  0   &  0   &  0   &  0   & 0\\
 0   &  0   &  0   &\w^{11}&  0   &  0   &  0   &  0   &  0   &  0   &  0   & 0\\
 0   &  0   &  0   &  0   &\w^{17}&  0   &  0   &  0   &  0   &  0   &  0   & 0\\
 0   &  0   &  0   &  0   &  0   &\w^{11}&  0   &  0   &  0   &  0   &  0   & 0
\end{array}
\right),\label{eq:Z12}
\end{align}
where $\omega_{24}$ denotes a primitive $24$th root of unity.

Like in the case of dimension $N=8$ discussed in detail in the
previous section, we consider a maximal Abelian subgroup of the
normalizer $\mathcal{N}_Z$ of the Zauner symmetry $F_Z$ in the
Clifford group when choosing the basis of the eigenspace of $F_Z$ of
dimension $5$.  We find the non-normalized orthogonal basis
$\{v_0,v_1,v_2,v_3,v_4\}$ with the vectors
\def\w{\omega_3}
\begin{align}
v_0&=(0,   0,   0,   0,   0,  0,  0, 1,  0,   1,  0,  1)\\
v_1&=(1,   1,   1,   1,   1,  1,  1, 0,  1,   0,  1,  0)\\
v_2&=(1,   1,\w^2,\w^2,  \w, \w, \w, 0,  1,   0,\w^2, 0)\\
v_3&=(0,   0,   0,   0,   0,  0,  0, 1,  0,\w^2,  0, \w)\\
v_4&=(1,\w^2,  \w,   1,\w^2, \w,  1, 0, \w,   0,\w^2, 0)
\end{align}

In the approach for computing a solution with Zauner's symmetry reported in
\cite{Grassl08}, computing a single modular Gr\"obner basis took about $40$
hours using more than $17$~GB of memory.  Using the $k$-nomial representation of
the Clifford group and an adapted basis of the eigenspace of Zauner's matrix,
the corresponding step uses less than $100$~MB of memory and takes less than
$40$ seconds on the same hardware--reduced to less than $25$ seconds on current
computers.  Moreover, instead of more than $300$ different primes, now only $21$
suffice. Overall, we get a speed-up by more than $3$ orders of magnitude.

A non-normalized fiducial vector
\begin{equation}\label{eq:fiducial12_sparse}
|\widetilde{\psi}_0\rangle=\sum_{i=0}^{11}x_i|b_i\rangle
\end{equation}
with respect to the representation of the Weyl-Heisenberg group generated by the
matrices $X$ and $Z$ given in Eqs.~(\ref{eq:X12}) and (\ref{eq:Z12}),
respectively, can be expressed in the number field
\begin{equation}
  {\mathbb{K}}={\mathbb{Q}}(\sqrt{2},\sqrt{3},\sqrt{13},t_1,s_1,s_2,\sqrt{-1})
\end{equation}
of degree $192$, where
\begin{equation}
s_1=\sqrt{(\sqrt{13}-1)/2}\,,\quad
s_2=\sqrt{(3\sqrt{13}+9)/2}\,,\quad\text{and}\quad
t_1^3=12 t_1+10.
\end{equation}
In this field, a primitive $24$th root of unity is given by
\begin{equation}
\omega_{24}=\frac{\sqrt{2}}{4}\left((1-\sqrt{3})+(\sqrt{3} + 1)\sqrt{-1}\right).
\end{equation}

The coefficients of $|\widetilde{\psi}_0\rangle$ in
Eq.~(\ref{eq:fiducial12_sparse}) are as follows:
\begin{footnotesize}
\begin{align}
x_{0}={}&    ((-30\sqrt{13}-312)s_1+(24\sqrt{13}t_1^2-102\sqrt{13}t_1-309\sqrt{13}-92t_1^2-158t_1-5))\sqrt{-1}\displaybreak[0]\nonumber\\
 & +(26\sqrt{39}-364\sqrt{3})s_1+28\sqrt{39}t_1^2-58\sqrt{39}t_1-147\sqrt{39}+96\sqrt{3}t_1^2-42\sqrt{3}t_1-443\sqrt{3}\displaybreak[3]\nonumber\\[1ex plus 3ex]
x_{1}={}&    (24\sqrt{13}t_1^2-102\sqrt{13}t_1-540\sqrt{13}-92t_1^2-158t_1-980)\sqrt{-1}\displaybreak[0]\nonumber\\
 & +28\sqrt{39}t_1^2-58\sqrt{39}t_1-264\sqrt{39}+96\sqrt{3}t_1^2-42\sqrt{3}t_1-1184\sqrt{3}\displaybreak[3]\nonumber\\[1ex plus 3ex]
x_{2}={}&    ((24\sqrt{13}-702)s_1-54\sqrt{13}t_1^2+138\sqrt{13}t_1+375\sqrt{13}-98t_1^2+142t_1+667)\sqrt{-1}\displaybreak[0]\nonumber\\
 & +(28\sqrt{39}-26\sqrt{3})s_1-2\sqrt{39}t_1^2-22\sqrt{39}t_1-81\sqrt{39}-94\sqrt{3}t_1^2-58\sqrt{3}t_1+219\sqrt{3}\displaybreak[3]\nonumber\\[1ex plus 3ex]
x_{3}={}&    ((-30\sqrt{13}-312)s_1-54\sqrt{13}t_1^2+138\sqrt{13}t_1+315\sqrt{13}-98t_1^2+142t_1+43)\sqrt{-1}\displaybreak[0]\nonumber\\
 & +(26\sqrt{39}-364\sqrt{3})s_1-2\sqrt{39}t_1^2-22\sqrt{39}t_1+93\sqrt{39}-94\sqrt{3}t_1^2-58\sqrt{3}t_1+1077\sqrt{3}\displaybreak[3]\nonumber\\[1ex plus 3ex]
x_{4}={}&    (30\sqrt{13}t_1^2-36\sqrt{13}t_1-588\sqrt{13}+190t_1^2+16t_1-3236)\sqrt{-1}\displaybreak[0]\nonumber\\
 &  -26\sqrt{39}t_1^2+80\sqrt{39}t_1+168\sqrt{39}-2\sqrt{3}t_1^2+100\sqrt{3}t_1-400\sqrt{3}\displaybreak[3]\nonumber\\[1ex plus 3ex]
x_{5}={}&    ((24\sqrt{13}-702)s_1+(30\sqrt{13}t_1^2-36\sqrt{13}t_1-297\sqrt{13}+190t_1^2+16t_1-1637))\sqrt{-1}\displaybreak[0]\nonumber\\
 & +(28\sqrt{39}-26\sqrt{3})s_1-26\sqrt{39}t_1^2+80\sqrt{39}t_1+111\sqrt{39}-2\sqrt{3}t_1^2+100\sqrt{3}t_1-517\sqrt{3}\displaybreak[3]\nonumber\\[1ex plus 3ex]
x_{6}={}&    ((-30\sqrt{13}-312)s_1+(30\sqrt{13}t_1^2-36\sqrt{13}t_1-357\sqrt{13}+190t_1^2+16t_1-2261))\sqrt{-1}\displaybreak[0]\nonumber\\
 & +(26\sqrt{39}-364\sqrt{3})s_1-26\sqrt{39}t_1^2+80\sqrt{39}t_1+285\sqrt{39}-2\sqrt{3}t_1^2+100\sqrt{3}t_1+341\sqrt{3}\displaybreak[3]\nonumber\\[1ex plus 3ex]
x_{7}={}&    488\sqrt{13}s_2\displaybreak[3]\nonumber\\[1ex plus 3ex]
x_{8}={}&    ((24\sqrt{13}-702)s_1+(24\sqrt{13}t_1^2-102\sqrt{13}t_1-249\sqrt{13}-92t_1^2-158t_1+619))\sqrt{-1}\displaybreak[0]\nonumber\\
 & +(28\sqrt{39}-26\sqrt{3})s_1+28\sqrt{39}t_1^2-58\sqrt{39}t_1-321\sqrt{39}+96\sqrt{3}t_1^2-42\sqrt{3}t_1-1301\sqrt{3}\displaybreak[3]\nonumber\\[1ex plus 3ex]
x_{9}={}&    ((85\sqrt{39}+91\sqrt{3})s_1s_2-122\sqrt{39}s_2)\sqrt{-1}\displaybreak[0]\nonumber\\
 & +(23\sqrt{13}-871)s_1s_2+122\sqrt{13}s_2\displaybreak[3]\nonumber\\[1ex plus 3ex]
x_{10}={}&   (-54\sqrt{13}t_1^2+138\sqrt{13}t_1+84\sqrt{13}-98t_1^2+142t_1-932)\sqrt{-1}\displaybreak[0]\nonumber\\
 &  -2\sqrt{39}t_1^2-22\sqrt{39}t_1-24\sqrt{39}-94\sqrt{3}t_1^2-58\sqrt{3}t_1+336\sqrt{3}\displaybreak[3]\nonumber\\[1ex plus 3ex]
x_{11}={}&   ((31\sqrt{39}+481\sqrt{3})s_1s_2+122\sqrt{39}s_2)\sqrt{-1}\displaybreak[0]\nonumber\\
 & +(139\sqrt{13}-299)s_1s_2+122\sqrt{13}s_2
\end{align}
\end{footnotesize}%
This expression for the fiducial vector is much more compact than the page
filling ones given in Ref.~\cite{Grassl08} and the supplementary material of
Ref.~\cite{Scott10}.

\section{Dimension $28$ fiducial}
\label{sec:dim28}
In order to compute a fiducial vector for dimension $N=28=2^2\times 7$ with the
help of the computer algebra system Magma \cite{magma}, we first
compute a basis such that both $X^{14}$ and $Z^{14}$ are
diagonal. It turns out that such a change of basis is given by the
matrix
\begin{equation}\label{eq:basechange28}
\arraycolsep0.75\arraycolsep
T=\frac{1}{\sqrt{2}}
\left(
\begin{array}{rr}
1 & 1\\
1 &-1\\
\end{array}
\right)
\otimes
\left(
\begin{array}{*{14}{c}}
0 & 1 & 0 & 0 & 0 & 0 & 0 & 0 & 0 & 0 & 0 & 0 & 0 & 0\\
0 & 0 & 0 & 1 & 0 & 0 & 0 & 0 & 0 & 0 & 0 & 0 & 0 & 0\\
0 & 0 & 0 & 0 & 0 & 1 & 0 & 0 & 0 & 0 & 0 & 0 & 0 & 0\\
0 & 0 & 0 & 0 & 0 & 0 & 0 & 1 & 0 & 0 & 0 & 0 & 0 & 0\\
0 & 0 & 0 & 0 & 0 & 0 & 0 & 0 & 0 & 1 & 0 & 0 & 0 & 0\\
0 & 0 & 0 & 0 & 0 & 0 & 0 & 0 & 0 & 0 & 0 & 1 & 0 & 0\\
0 & 0 & 0 & 0 & 0 & 0 & 0 & 0 & 0 & 0 & 0 & 0 & 0 & 1\\
1 & 0 & 0 & 0 & 0 & 0 & 0 & 0 & 0 & 0 & 0 & 0 & 0 & 0\\
0 & 0 & 1 & 0 & 0 & 0 & 0 & 0 & 0 & 0 & 0 & 0 & 0 & 0\\
0 & 0 & 0 & 0 & 1 & 0 & 0 & 0 & 0 & 0 & 0 & 0 & 0 & 0\\
0 & 0 & 0 & 0 & 0 & 0 & 1 & 0 & 0 & 0 & 0 & 0 & 0 & 0\\
0 & 0 & 0 & 0 & 0 & 0 & 0 & 0 & 1 & 0 & 0 & 0 & 0 & 0\\
0 & 0 & 0 & 0 & 0 & 0 & 0 & 0 & 0 & 0 & 1 & 0 & 0 & 0\\
0 & 0 & 0 & 0 & 0 & 0 & 0 & 0 & 0 & 0 & 0 & 0 & 1 & 0
\end{array}
\right),
\end{equation}
which is a tensor product of a $2\times 2$ Fourier transformation and
a permutation re-ordering the basis elements with odd/even index.  The
numerical solution in the orbit $28c$ in \cite{Scott10} indicates that
there is a solution which possesses an order-two anti-unitary symmetry
\begin{equation}
F_c=
\begin{pmatrix}
11&6\\
50&17
\end{pmatrix},
\end{equation}
in addition to the symmetry $F_z$ of order three conjectured by
Zauner.  In order to compute a basis of the corresponding eigenspace,
we represent the anti-unitary transformation $U_{F_cF_z}$ as a real
orthogonal matrix $O_{F_cF_z}\in O(2N)$ using the isomorphism
$\mathbb{C}\to\mathbb{R}^2$.  We obtain ten linearly independent
eigenvectors $\mathbf{v}'_i\in\mathbb{R}^{56}$ which are then mapped
back to ten vectors $|v_i\rangle\in\mathbb{C}^{28}$.  Then we write
the fiducial vector as
\begin{equation}\label{eq:fiducial28_eigenspace}
|\psi_0\rangle=\sum_{i=0}^9 y_i |v_i\rangle,
\end{equation}
with real variables $y_i$, $i=0,\ldots,9$.  The defining equations
for the SIC are given as
\begin{align}
\langle \psi_0|\psi_0\rangle&=1\qquad\text{and}\\
|\langle
\psi_0|D_{\mathbf{p}}|\psi_0\rangle|^2&=\frac{1}{29},\qquad\text{for $\mathbf{p}\ne 0$.}
\end{align}
We solved these equations via computing a Gr\"obner basis for the
corresponding polynomial ideal, using similar techniques to those
described in \cite{Grassl08}.  First, the equations are reformulated
as equations over the rationals, adding an auxiliary variable. This
allows us to compute a modular GB, i.e., performing all
computations modulo a large prime.

Using Magma V2.18 on a Linux PC with 3 GHz clock speed and 32 GB RAM,
the first step took about 60 hours and used up to 45 GB of memory.
One of the polynomials in the modular Gr\"obner basis could be lifted to the
rationals.  Adding this polynomial, the computation for three more
primes took only approximately six hours. Having computed four modular
Gr\"obner bases, six polynomials could be lifted to the rationals.  Adding those
polynomials reduced the time to compute a modular Gr\"obner basis to only five
minutes.  Then, with five different primes, we lifted $20$ polynomials
to the rationals.  The following Gr\"obner basis computations took less than one
second each.  In total we used $18$ different primes with about 23
bits each to obtain a Gr\"obner basis over the rationals.  As a next
step, the Gr\"obner basis is converted to a Gr\"obner basis with
respect to lexicographic order.  This yields a polynomial system of
equations in triangular form.  The coefficients of the polynomials
have numerators and denominators with more than 50 digits.  It turns
out that the ideal is zero-dimensional, i.e., there are finitely many
solutions.  Note that we are only interested in solutions where all
variables $y_i$ assume real values

Inserting one of these solution into
Eq.~(\ref{eq:fiducial28_eigenspace}), we find that a non-normalized
fiducial vector
\begin{equation}\label{eq:fiducial28_sparse}
|\widetilde{\psi}_0\rangle=\sum_{i=0}^{27}x_i|b_i\rangle
\end{equation}
in the adapted bases given by the matrix $T$ in
Eq.~(\ref{eq:basechange28}) can be expressed in the number field
\begin{equation}
  {\mathbb{K}}={\mathbb{Q}}(\sqrt{2},\sqrt{7},\sqrt{29},r_1,r_2,s_1,s_2,\sqrt{-1})
\end{equation}
of degree $576$, where
\begin{equation}
r_1=\sqrt{\sqrt{29}+5}\,,\quad
r_2=\sqrt{\sqrt{29}+1}\,,\quad
s_1^3=21 s_1-7\,,\quad\text{and}\quad
s_2^3=14 s_2-14.
\end{equation}
Note that even though we do not explicitly use a $56$th root of unity,
it can be expressed as
\begin{equation}
  \omega_{56}=\frac{1}{252}\Bigl((s_1^2-7s_1-35)(1+\sqrt{-1})\sqrt{14}-(7s_1^2-7s_1-77)(1-\sqrt{-1})\sqrt{2}\Bigr).
\end{equation}
The Galois group of ${\mathbb{K}}$ is isomorphic to
$C_6\times((C_6 \times C_2 \times C_2 \times C_2)\rtimes C_2)$, and
$\mathbb{K}$ is an Abelian extension of ${\mathbb{Q}}(\sqrt{29})$.

The coefficients of $|\widetilde{\psi}_0\rangle$ in
Eq.~(\ref{eq:fiducial28_sparse}) are as follows:
\begin{footnotesize}
\begin{align}
 x_{0}={}&(((2s_2s_1^2+(4s_2-21)s_1-7s_2+42)\sqrt{29}\displaybreak[0]\nonumber\\
        &+((-12s_2^2-18s_2+112)s_1^2+203s_1+(210s_2^2+315s_2-2366)))\sqrt{2}r_2\displaybreak[0]\nonumber\\
        &+((1/7(4s_2^2+28s_2)s_1^2+(8s_2^2+20s_2-84)s_1-26s_2^2-110s_2+168)\sqrt{29}\displaybreak[0]\nonumber\\
        &+(1/7(60s_2^2-84s_2-560)s_1^2+(-24s_2^2-36s_2+224)s_1-102s_2^2+282s_2+952))\sqrt{7})r_1\sqrt{-1}\displaybreak[0]\nonumber\\
        &+(((9s_1+9s_2)\sqrt{29}-87s_1+18s_2^2+27s_2-168)\sqrt{14}r_2\displaybreak[0]\nonumber\\
        &+((-84s_1+(18s_2^2+126s_2-336))\sqrt{29}+(270s_2^2-378s_2-2520)))r_1\displaybreak[3]\nonumber\\[1ex plus 3ex]
 x_{1}={}&(((-s_2s_1^2+(-2s_2+21)s_1+(35s_2+84))\sqrt{29}\displaybreak[0]\nonumber\\
        &+((-6s_2^2-9s_2+56)s_1^2-203s_1+42s_2^2+63s_2-1204))\sqrt{2}r_2\displaybreak[0]\nonumber\\
        &+((1/7(10s_2^2+28s_2-168)s_1^2+(8s_2^2+26s_2-84)s_1-14s_2^2-68s_2+336)\sqrt{29}\displaybreak[0]\nonumber\\
        &+(1/7(-18s_2^2-56s_2+168)s_1^2-58s_2s_1-90s_2^2+184s_2+840))\sqrt{7})r_1\sqrt{-1}\displaybreak[0]\nonumber\\
        &+((((s_2-2)s_1^2+(2s_2-1)s_1-17s_2+28)\sqrt{29}\displaybreak[0]\nonumber\\
        &+((-2s_2^2-3s_2+38)s_1^2+(8s_2^2+12s_2-65)s_1+(10s_2^2+15s_2-364)))\sqrt{14}r_2\displaybreak[0]\nonumber\\
        &+(((-2s_2^2-8s_2)s_1^2+(8s_2^2+14s_2-84)s_1+(46s_2^2+112s_2-336))\sqrt{29}\displaybreak[0]\nonumber\\
        &+((-6s_2^2+20s_2+56)s_1^2+(-48s_2^2-14s_2+448)s_1-150s_2^2-196s_2+1400)))r_1\displaybreak[3]\nonumber\\[1ex plus 3ex]
 x_{2}={}&((2s_2s_1^2-8s_2s_1-28s_2)\sqrt{58}r_2\displaybreak[0]\nonumber\\
        &+((1/7(8s_2^2+14s_2-112)s_1^2+(4s_2^2+16s_2-56)s_1-28s_2^2-40s_2+392)\sqrt{29}\displaybreak[0]\nonumber\\
        &+(1/7(-48s_2^2-14s_2+448)s_1^2+(12s_2^2-40s_2-112)s_1+(204s_2^2+16s_2-1904)))\sqrt{7})r_1\sqrt{-1}\displaybreak[0]\nonumber\\
        &+(((-2s_1^2-4s_1+70)\sqrt{29}\displaybreak[0]\nonumber\\
        &+((4s_2^2+6s_2-18)s_1^2+(8s_2^2+12s_2-36)s_1-32s_2^2-48s_2-378))\sqrt{14}r_2\displaybreak[0]\nonumber\\
        &+(((-4s_2^2-10s_2+56)s_1^2+(4s_2^2+28s_2-56)s_1+(116s_2^2+308s_2-1120))\sqrt{29}\displaybreak[0]\nonumber\\
        &+((12s_2^2+18s_2-112)s_1^2+(60s_2^2-84s_2-560)s_1-276s_2^2-588s_2+2576)))r_1\displaybreak[3]\nonumber\\[1ex plus 3ex]
 x_{3}={}&((1/7(-16s_2^2-28s_2+168)s_1^2+(4s_2^2+16s_2)s_1+(68s_2^2+128s_2-672))\sqrt{29}\displaybreak[0]\nonumber\\
        &+(1/7(96s_2^2+28s_2-896)s_1^2+(12s_2^2-40s_2-112)s_1-372s_2^2-152s_2+3472))\sqrt{7}r_1\sqrt{-1}\displaybreak[0]\nonumber\\
        &+(((-8s_2^2-20s_2+56)s_1^2+(-52s_2^2-112s_2+448)s_1+(172s_2^2+448s_2-1120))\sqrt{29}\displaybreak[0]\nonumber\\
        &+((24s_2^2+36s_2-224)s_1^2+(228s_2^2+168s_2-2128)s_1-444s_2^2-840s_2+4144))r_1\displaybreak[3]\nonumber\\[1ex plus 3ex]
 x_{4}={}&((-2s_2s_1^2+8s_2s_1+28s_2)\sqrt{58}r_2\displaybreak[0]\nonumber\\
        &+((1/7(8s_2^2+14s_2-112)s_1^2+(4s_2^2+16s_2-56)s_1-28s_2^2-40s_2+392)\sqrt{29}\displaybreak[0]\nonumber\\
        &+(1/7(-48s_2^2-14s_2+448)s_1^2+(12s_2^2-40s_2-112)s_1+(204s_2^2+16s_2-1904)))\sqrt{7})r_1\sqrt{-1}\displaybreak[0]\nonumber\\
        &+(((2s_1^2+4s_1-70)\sqrt{29}\displaybreak[0]\nonumber\\
        &+((-4s_2^2-6s_2+18)s_1^2+(-8s_2^2-12s_2+36)s_1+(32s_2^2+48s_2+378)))\sqrt{14}r_2\displaybreak[0]\nonumber\\
        &+(((-4s_2^2-10s_2+56)s_1^2+(4s_2^2+28s_2-56)s_1+(116s_2^2+308s_2-1120))\sqrt{29}\displaybreak[0]\nonumber\\
        &+((12s_2^2+18s_2-112)s_1^2+(60s_2^2-84s_2-560)s_1-276s_2^2-588s_2+2576)))r_1\displaybreak[3]\nonumber\\[1ex plus 3ex]
 x_{5}={}&(((s_2s_1^2+(2s_2-21)s_1-35s_2-84)\sqrt{29}\displaybreak[0]\nonumber\\
        &+((6s_2^2+9s_2-56)s_1^2+203s_1-42s_2^2-63s_2+1204))\sqrt{2}r_2\displaybreak[0]\nonumber\\
        &+((1/7(10s_2^2+28s_2-168)s_1^2+(8s_2^2+26s_2-84)s_1-14s_2^2-68s_2+336)\sqrt{29}\displaybreak[0]\nonumber\\
        &+(1/7(-18s_2^2-56s_2+168)s_1^2-58s_2s_1-90s_2^2+184s_2+840))\sqrt{7})r_1\sqrt{-1}\displaybreak[0]\nonumber\\
        &+((((-s_2+2)s_1^2+(-2s_2+1)s_1+(17s_2-28))\sqrt{29}\displaybreak[0]\nonumber\\
        &+((2s_2^2+3s_2-38)s_1^2+(-8s_2^2-12s_2+65)s_1-10s_2^2-15s_2+364))\sqrt{14}r_2\displaybreak[0]\nonumber\\
        &+(((-2s_2^2-8s_2)s_1^2+(8s_2^2+14s_2-84)s_1+(46s_2^2+112s_2-336))\sqrt{29}\displaybreak[0]\nonumber\\
        &+((-6s_2^2+20s_2+56)s_1^2+(-48s_2^2-14s_2+448)s_1-150s_2^2-196s_2+1400)))r_1\displaybreak[3]\nonumber\\[1ex plus 3ex]
 x_{6}={}&(((-2s_2s_1^2+(-4s_2+21)s_1+(7s_2-42))\sqrt{29}\displaybreak[0]\nonumber\\
        &+((12s_2^2+18s_2-112)s_1^2-203s_1-210s_2^2-315s_2+2366))\sqrt{2}r_2\displaybreak[0]\nonumber\\
        &+((1/7(4s_2^2+28s_2)s_1^2+(8s_2^2+20s_2-84)s_1-26s_2^2-110s_2+168)\sqrt{29}\displaybreak[0]\nonumber\\
        &+(1/7(60s_2^2-84s_2-560)s_1^2+(-24s_2^2-36s_2+224)s_1-102s_2^2+282s_2+952))\sqrt{7})r_1\sqrt{-1}\displaybreak[0]\nonumber\\
        &+(((-9s_1-9s_2)\sqrt{29}+(87s_1-18s_2^2-27s_2+168))\sqrt{14}r_2\displaybreak[0]\nonumber\\
        &+((-84s_1+(18s_2^2+126s_2-336))\sqrt{29}+(270s_2^2-378s_2-2520)))r_1\displaybreak[3]\nonumber\\[1ex plus 3ex]
 x_{7}={}&(((1/7(10s_2^2+28s_2-84)s_1^2+(2s_2^2+2s_2)s_1-38s_2^2-110s_2+336)\sqrt{29}\displaybreak[0]\nonumber\\
        &+(1/7(-18s_2^2-56s_2+168)s_1^2+(-18s_2^2+2s_2+168)s_1+(54s_2^2+226s_2-504)))\sqrt{7}\displaybreak[0]\nonumber\\
        &+(((-2s_2^2-8s_2+28)s_1^2+(-22s_2^2-70s_2+224)s_1+(34s_2^2+154s_2-560))\sqrt{29}\displaybreak[0]\nonumber\\
        &+((-6s_2^2+20s_2+56)s_1^2+(6s_2^2+154s_2-56)s_1+(174s_2^2-406s_2-1624))))\sqrt{2}r_1\sqrt{-1}\displaybreak[0]\nonumber\\
        &+(((1/7(10s_2^2+28s_2-84)s_1^2+(2s_2^2+2s_2)s_1-38s_2^2-110s_2+336)\sqrt{29}\displaybreak[0]\nonumber\\
        &+(1/7(-18s_2^2-56s_2+168)s_1^2+(-18s_2^2+2s_2+168)s_1+(54s_2^2+226s_2-504)))\sqrt{7}\displaybreak[0]\nonumber\\
        &+(((2s_2^2+8s_2-28)s_1^2+(22s_2^2+70s_2-224)s_1-34s_2^2-154s_2+560)\sqrt{29}\displaybreak[0]\nonumber\\
        &+((6s_2^2-20s_2-56)s_1^2+(-6s_2^2-154s_2+56)s_1-174s_2^2+406s_2+1624)))\sqrt{2}r_1\displaybreak[3]\nonumber\\[1ex plus 3ex]
 x_{8}={}&(((((s_2-2)s_1^2+(2s_2-4)s_1-8s_2+70)\sqrt{29}\displaybreak[0]\nonumber\\
        &+((-2s_2^2-3s_2+38)s_1^2+(-4s_2^2-6s_2+76)s_1+(16s_2^2+24s_2-826)))\sqrt{7}\displaybreak[0]\nonumber\\
        &+((s_2s_1^2-4s_2s_1-14s_2)\sqrt{29}\displaybreak[0]\nonumber\\
        &+((6s_2^2+9s_2-56)s_1^2+(-24s_2^2-36s_2+224)s_1-84s_2^2-126s_2+784)))r_2\displaybreak[0]\nonumber\\
        &+(((1/7(-5s_2^2-14s_2+28)s_1^2+(2s_2^2+2s_2-28)s_1+(22s_2^2+58s_2-140))\sqrt{29}\displaybreak[0]\nonumber\\
        &+(1/7(9s_2^2+28s_2-84)s_1^2+(-18s_2^2+2s_2+168)s_1-54s_2^2-110s_2+504))\sqrt{7}\displaybreak[0]\nonumber\\
        &+(((-s_2^2-4s_2)s_1^2+(-8s_2^2-14s_2+84)s_1+(20s_2^2+98s_2-168))\sqrt{29}\displaybreak[0]\nonumber\\
        &+((-3s_2^2+10s_2+28)s_1^2+(48s_2^2+14s_2-448)s_1+(132s_2^2-266s_2-1232))))\sqrt{2})r_1\sqrt{-1}\displaybreak[0]\nonumber\\
        &+(((((-s_2+2)s_1^2+(-2s_2+4)s_1+(8s_2-70))\sqrt{29}\displaybreak[0]\nonumber\\
        &+((2s_2^2+3s_2-38)s_1^2+(4s_2^2+6s_2-76)s_1-16s_2^2-24s_2+826))\sqrt{7}\displaybreak[0]\nonumber\\
        &+((s_2s_1^2-4s_2s_1-14s_2)\sqrt{29}\displaybreak[0]\nonumber\\
        &+((6s_2^2+9s_2-56)s_1^2+(-24s_2^2-36s_2+224)s_1-84s_2^2-126s_2+784)))r_2\displaybreak[0]\nonumber\\
        &+(((1/7(-5s_2^2-14s_2+28)s_1^2+(2s_2^2+2s_2-28)s_1+(22s_2^2+58s_2-140))\sqrt{29}\displaybreak[0]\nonumber\\
        &+(1/7(9s_2^2+28s_2-84)s_1^2+(-18s_2^2+2s_2+168)s_1-54s_2^2-110s_2+504))\sqrt{7}\displaybreak[0]\nonumber\\
        &+(((s_2^2+4s_2)s_1^2+(8s_2^2+14s_2-84)s_1-20s_2^2-98s_2+168)\sqrt{29}\displaybreak[0]\nonumber\\
        &+((3s_2^2-10s_2-28)s_1^2+(-48s_2^2-14s_2+448)s_1-132s_2^2+266s_2+1232)))\sqrt{2})r_1\displaybreak[3]\nonumber\\[1ex plus 3ex]
 x_{9}={}&(((((-s_2-2)s_1^2+(s_2-1)s_1+(11s_2+28))\sqrt{29}\displaybreak[0]\nonumber\\
        &+((-2s_2^2-3s_2+38)s_1^2+(-10s_2^2-15s_2+103)s_1+(46s_2^2+69s_2-700)))\sqrt{7}\displaybreak[0]\nonumber\\
        &+((s_2s_1^2+(-s_2-21)s_1+(7s_2-84))\sqrt{29}\displaybreak[0]\nonumber\\
        &+((-6s_2^2-9s_2+56)s_1^2+(-6s_2^2-9s_2+259)s_1+(126s_2^2+189s_2-364))))r_2\displaybreak[0]\nonumber\\
        &+(((1/7(-s_2^2-7s_2-28)s_1^2+(s_2^2-2s_2-14)s_1-13s_2^2-19s_2+224)\sqrt{29}\displaybreak[0]\nonumber\\
        &+(1/7(-15s_2^2+21s_2+140)s_1^2+(-21s_2^2+12s_2+196)s_1+(93s_2^2+9s_2-868)))\sqrt{7}\displaybreak[0]\nonumber\\
        &+(((s_2^2+s_2)s_1^2+(5s_2^2+14s_2-42)s_1+(13s_2^2+49s_2-168))\sqrt{29}\displaybreak[0]\nonumber\\
        &+((-9s_2^2+s_2+84)s_1^2+(-9s_2^2-28s_2+84)s_1+(27s_2^2-119s_2-252))))\sqrt{2})r_1\sqrt{-1}\displaybreak[0]\nonumber\\
        &+(((((s_2+2)s_1^2+(-s_2+1)s_1-11s_2-28)\sqrt{29}\displaybreak[0]\nonumber\\
        &+((2s_2^2+3s_2-38)s_1^2+(10s_2^2+15s_2-103)s_1-46s_2^2-69s_2+700))\sqrt{7}\displaybreak[0]\nonumber\\
        &+((s_2s_1^2+(-s_2-21)s_1+(7s_2-84))\sqrt{29}\displaybreak[0]\nonumber\\
        &+((-6s_2^2-9s_2+56)s_1^2+(-6s_2^2-9s_2+259)s_1+(126s_2^2+189s_2-364))))r_2\displaybreak[0]\nonumber\\
        &+(((1/7(-s_2^2-7s_2-28)s_1^2+(s_2^2-2s_2-14)s_1-13s_2^2-19s_2+224)\sqrt{29}\displaybreak[0]\nonumber\\
        &+(1/7(-15s_2^2+21s_2+140)s_1^2+(-21s_2^2+12s_2+196)s_1+(93s_2^2+9s_2-868)))\sqrt{7}\displaybreak[0]\nonumber\\
        &+(((-s_2^2-s_2)s_1^2+(-5s_2^2-14s_2+42)s_1-13s_2^2-49s_2+168)\sqrt{29}\displaybreak[0]\nonumber\\
        &+((9s_2^2-s_2-84)s_1^2+(9s_2^2+28s_2-84)s_1-27s_2^2+119s_2+252)))\sqrt{2})r_1\displaybreak[3]\nonumber\\[1ex plus 3ex]
x_{10}={}&(((9s_1\sqrt{29}-87s_1-36s_2^2-54s_2+336)\sqrt{7}+((4s_2s_1^2+(2s_2+21)s_1-56s_2-42)\sqrt{29}\displaybreak[0]\nonumber\\
        &+((12s_2^2+18s_2-315)s_1+(84s_2^2+126s_2-378))))r_2\displaybreak[0]\nonumber\\
        &+(((1/7(8s_2^2+14s_2-56)s_1^2+(-2s_2^2-8s_2+14)s_1-16s_2^2-19s_2+112)\sqrt{29}\displaybreak[0]\nonumber\\
        &+(1/7(-48s_2^2-14s_2+448)s_1^2+(-6s_2^2+20s_2+56)s_1+(132s_2^2-5s_2-1232)))\sqrt{7}\displaybreak[0]\nonumber\\
        &+((42s_1-36s_2^2-63s_2+420)\sqrt{29}+(216s_2^2+63s_2-2016)))\sqrt{2})r_1\sqrt{-1}\displaybreak[0]\nonumber\\
        &+(((-9s_1\sqrt{29}+(87s_1+(36s_2^2+54s_2-336)))\sqrt{7}+((4s_2s_1^2+(2s_2+21)s_1-56s_2-42)\sqrt{29}\displaybreak[0]\nonumber\\
        &+((12s_2^2+18s_2-315)s_1+(84s_2^2+126s_2-378))))r_2\displaybreak[0]\nonumber\\
        &+(((1/7(8s_2^2+14s_2-56)s_1^2+(-2s_2^2-8s_2+14)s_1-16s_2^2-19s_2+112)\sqrt{29}\displaybreak[0]\nonumber\\
        &+(1/7(-48s_2^2-14s_2+448)s_1^2+(-6s_2^2+20s_2+56)s_1+(132s_2^2-5s_2-1232)))\sqrt{7}\displaybreak[0]\nonumber\\
        &+((-42s_1+(36s_2^2+63s_2-420))\sqrt{29}+(-216s_2^2-63s_2+2016)))\sqrt{2})r_1\displaybreak[3]\nonumber\\[1ex plus 3ex]
x_{11}={}&(((-9s_1\sqrt{29}+(87s_1+(36s_2^2+54s_2-336)))\sqrt{7}+((-4s_2s_1^2+(-2s_2-21)s_1+(56s_2+42))\sqrt{29}\displaybreak[0]\nonumber\\
        &+((-12s_2^2-18s_2+315)s_1-84s_2^2-126s_2+378)))r_2\displaybreak[0]\nonumber\\
        &+(((1/7(8s_2^2+14s_2-56)s_1^2+(-2s_2^2-8s_2+14)s_1-16s_2^2-19s_2+112)\sqrt{29}\displaybreak[0]\nonumber\\
        &+(1/7(-48s_2^2-14s_2+448)s_1^2+(-6s_2^2+20s_2+56)s_1+(132s_2^2-5s_2-1232)))\sqrt{7}\displaybreak[0]\nonumber\\
        &+((42s_1-36s_2^2-63s_2+420)\sqrt{29}+(216s_2^2+63s_2-2016)))\sqrt{2})r_1\sqrt{-1}\displaybreak[0]\nonumber\\
        &+(((9s_1\sqrt{29}-87s_1-36s_2^2-54s_2+336)\sqrt{7}+((-4s_2s_1^2+(-2s_2-21)s_1+(56s_2+42))\sqrt{29}\displaybreak[0]\nonumber\\
        &+((-12s_2^2-18s_2+315)s_1-84s_2^2-126s_2+378)))r_2\displaybreak[0]\nonumber\\
        &+(((1/7(8s_2^2+14s_2-56)s_1^2+(-2s_2^2-8s_2+14)s_1-16s_2^2-19s_2+112)\sqrt{29}\displaybreak[0]\nonumber\\
        &+(1/7(-48s_2^2-14s_2+448)s_1^2+(-6s_2^2+20s_2+56)s_1+(132s_2^2-5s_2-1232)))\sqrt{7}\displaybreak[0]\nonumber\\
        &+((-42s_1+(36s_2^2+63s_2-420))\sqrt{29}+(-216s_2^2-63s_2+2016)))\sqrt{2})r_1\displaybreak[3]\nonumber\\[1ex plus 3ex]
x_{12}={}&(((((s_2+2)s_1^2+(-s_2+1)s_1-11s_2-28)\sqrt{29}\displaybreak[0]\nonumber\\
        &+((2s_2^2+3s_2-38)s_1^2+(10s_2^2+15s_2-103)s_1-46s_2^2-69s_2+700))\sqrt{7}\displaybreak[0]\nonumber\\
        &+((-s_2s_1^2+(s_2+21)s_1-7s_2+84)\sqrt{29}\displaybreak[0]\nonumber\\
        &+((6s_2^2+9s_2-56)s_1^2+(6s_2^2+9s_2-259)s_1-126s_2^2-189s_2+364)))r_2\displaybreak[0]\nonumber\\
        &+(((1/7(-s_2^2-7s_2-28)s_1^2+(s_2^2-2s_2-14)s_1-13s_2^2-19s_2+224)\sqrt{29}\displaybreak[0]\nonumber\\
        &+(1/7(-15s_2^2+21s_2+140)s_1^2+(-21s_2^2+12s_2+196)s_1+(93s_2^2+9s_2-868)))\sqrt{7}\displaybreak[0]\nonumber\\
        &+(((s_2^2+s_2)s_1^2+(5s_2^2+14s_2-42)s_1+(13s_2^2+49s_2-168))\sqrt{29}\displaybreak[0]\nonumber\\
        &+((-9s_2^2+s_2+84)s_1^2+(-9s_2^2-28s_2+84)s_1+(27s_2^2-119s_2-252))))\sqrt{2})r_1\sqrt{-1}\displaybreak[0]\nonumber\\
        &+(((((-s_2-2)s_1^2+(s_2-1)s_1+(11s_2+28))\sqrt{29}\displaybreak[0]\nonumber\\
        &+((-2s_2^2-3s_2+38)s_1^2+(-10s_2^2-15s_2+103)s_1+(46s_2^2+69s_2-700)))\sqrt{7}\displaybreak[0]\nonumber\\
        &+((-s_2s_1^2+(s_2+21)s_1-7s_2+84)\sqrt{29}\displaybreak[0]\nonumber\\
        &+((6s_2^2+9s_2-56)s_1^2+(6s_2^2+9s_2-259)s_1-126s_2^2-189s_2+364)))r_2\displaybreak[0]\nonumber\\
        &+(((1/7(-s_2^2-7s_2-28)s_1^2+(s_2^2-2s_2-14)s_1-13s_2^2-19s_2+224)\sqrt{29}\displaybreak[0]\nonumber\\
        &+(1/7(-15s_2^2+21s_2+140)s_1^2+(-21s_2^2+12s_2+196)s_1+(93s_2^2+9s_2-868)))\sqrt{7}\displaybreak[0]\nonumber\\
        &+(((-s_2^2-s_2)s_1^2+(-5s_2^2-14s_2+42)s_1-13s_2^2-49s_2+168)\sqrt{29}\displaybreak[0]\nonumber\\
        &+((9s_2^2-s_2-84)s_1^2+(9s_2^2+28s_2-84)s_1-27s_2^2+119s_2+252)))\sqrt{2})r_1\displaybreak[3]\nonumber\\[1ex plus 3ex]
x_{13}={}&(((((-s_2+2)s_1^2+(-2s_2+4)s_1+(8s_2-70))\sqrt{29}\displaybreak[0]\nonumber\\
        &+((2s_2^2+3s_2-38)s_1^2+(4s_2^2+6s_2-76)s_1-16s_2^2-24s_2+826))\sqrt{7}+((-s_2s_1^2+4s_2s_1+14s_2)\sqrt{29}\displaybreak[0]\nonumber\\
        &+((-6s_2^2-9s_2+56)s_1^2+(24s_2^2+36s_2-224)s_1+(84s_2^2+126s_2-784))))r_2\displaybreak[0]\nonumber\\
        &+(((1/7(-5s_2^2-14s_2+28)s_1^2+(2s_2^2+2s_2-28)s_1+(22s_2^2+58s_2-140))\sqrt{29}\displaybreak[0]\nonumber\\
        &+(1/7(9s_2^2+28s_2-84)s_1^2+(-18s_2^2+2s_2+168)s_1-54s_2^2-110s_2+504))\sqrt{7}\displaybreak[0]\nonumber\\
        &+(((-s_2^2-4s_2)s_1^2+(-8s_2^2-14s_2+84)s_1+(20s_2^2+98s_2-168))\sqrt{29}\displaybreak[0]\nonumber\\
        &+((-3s_2^2+10s_2+28)s_1^2+(48s_2^2+14s_2-448)s_1+(132s_2^2-266s_2-1232))))\sqrt{2})r_1\sqrt{-1}\displaybreak[0]\nonumber\\
        &+(((((s_2-2)s_1^2+(2s_2-4)s_1-8s_2+70)\sqrt{29}+((-2s_2^2-3s_2+38)s_1^2\displaybreak[0]\nonumber\\
        &+(-4s_2^2-6s_2+76)s_1+(16s_2^2+24s_2-826)))\sqrt{7}+((-s_2s_1^2+4s_2s_1+14s_2)\sqrt{29}\displaybreak[0]\nonumber\\
        &+((-6s_2^2-9s_2+56)s_1^2+(24s_2^2+36s_2-224)s_1+(84s_2^2+126s_2-784))))r_2\displaybreak[0]\nonumber\\
        &+(((1/7(-5s_2^2-14s_2+28)s_1^2+(2s_2^2+2s_2-28)s_1+(22s_2^2+58s_2-140))\sqrt{29}\displaybreak[0]\nonumber\\
        &+(1/7(9s_2^2+28s_2-84)s_1^2+(-18s_2^2+2s_2+168)s_1-54s_2^2-110s_2+504))\sqrt{7}\displaybreak[0]\nonumber\\
        &+(((s_2^2+4s_2)s_1^2+(8s_2^2+14s_2-84)s_1-20s_2^2-98s_2+168)\sqrt{29}\displaybreak[0]\nonumber\\
        &+((3s_2^2-10s_2-28)s_1^2+(-48s_2^2-14s_2+448)s_1-132s_2^2+266s_2+1232)))\sqrt{2})r_1\displaybreak[3]\nonumber\\[1ex plus 3ex]
x_{14}={}&((-18s_1\sqrt{203}+(42s_1-84)\sqrt{29})\sqrt{2}r_2+(((16s_1^2-34s_1-248)\sqrt{29}-174s_1-696)\sqrt{7}\displaybreak[0]\nonumber\\
        &+((-8s_1^2-322s_1-476)\sqrt{29}-232s_1^2-2030s_1+812)))\sqrt{-1}\displaybreak[0]\nonumber\\
        &+(-18s_1\sqrt{203}+(-42s_1+84)\sqrt{29})\sqrt{2}r_2+((-16s_1^2+34s_1+248)\sqrt{29}+(174s_1+696))\sqrt{7}\displaybreak[0]\nonumber\\
        &+(-8s_1^2-322s_1-476)\sqrt{29}-232s_1^2-2030s_1+812\displaybreak[3]\nonumber\\[1ex plus 3ex]
x_{15}={}&(((-4s_1^2-2s_1+56)\sqrt{203}+(42s_1+168)\sqrt{29})\sqrt{2}r_2\displaybreak[0]\nonumber\\
        &+(((40s_1^2+26s_1-788)\sqrt{29}+(232s_1^2-58s_1-4988))\sqrt{7}\displaybreak[0]\nonumber\\
        &+((60s_1^2+42s_1-252)\sqrt{29}+(116s_1^2+406s_1+812))))\sqrt{-1}\displaybreak[0]\nonumber\\
        &+((-4s_1^2-2s_1+56)\sqrt{203}+(-42s_1-168)\sqrt{29})\sqrt{2}r_2\displaybreak[0]\nonumber\\
        &+((-40s_1^2-26s_1+788)\sqrt{29}-232s_1^2+58s_1+4988)\sqrt{7}\displaybreak[0]\nonumber\\
        &+(60s_1^2+42s_1-252)\sqrt{29}+116s_1^2+406s_1+812\displaybreak[3]\nonumber\\[1ex plus 3ex]
x_{16}={}&((4s_1^2+8s_1-140)\sqrt{406}r_2+(((-16s_1^2-68s_1+452)\sqrt{29}-348s_1+1740)\sqrt{7}+((60s_1^2-84s_1)\sqrt{29}\displaybreak[0]\nonumber\\
        &+(116s_1^2-812s_1+3248))))\sqrt{-1}+(4s_1^2+8s_1-140)\sqrt{406}r_2\displaybreak[0]\nonumber\\
        &+((16s_1^2+68s_1-452)\sqrt{29}+(348s_1-1740))\sqrt{7}+(60s_1^2-84s_1)\sqrt{29}+116s_1^2-812s_1+3248\displaybreak[3]\nonumber\\[1ex plus 3ex]
x_{17}={}&(((-12s_1^2-144s_1+192)\sqrt{29}-116s_1^2-928s_1+2320)\sqrt{7}+((68s_1^2+112s_1-1792)\sqrt{29}\displaybreak[0]\nonumber\\
        &+(348s_1^2-9744)))\sqrt{-1}+((12s_1^2+144s_1-192)\sqrt{29}+(116s_1^2+928s_1-2320))\sqrt{7}\displaybreak[0]\nonumber\\
        &+(68s_1^2+112s_1-1792)\sqrt{29}+348s_1^2-9744\displaybreak[3]\nonumber\\[1ex plus 3ex]
x_{18}={}&((-4s_1^2-8s_1+140)\sqrt{406}r_2+(((-16s_1^2-68s_1+452)\sqrt{29}-348s_1+1740)\sqrt{7}\displaybreak[0]\nonumber\\
        &+((60s_1^2-84s_1)\sqrt{29}+(116s_1^2-812s_1+3248))))\sqrt{-1}\displaybreak[0]\nonumber\\
        &+(-4s_1^2-8s_1+140)\sqrt{406}r_2+((16s_1^2+68s_1-452)\sqrt{29}+(348s_1-1740))\sqrt{7}\displaybreak[0]\nonumber\\
        &+(60s_1^2-84s_1)\sqrt{29}+116s_1^2-812s_1+3248\displaybreak[3]\nonumber\\[1ex plus 3ex]
x_{19}={}&(((4s_1^2+2s_1-56)\sqrt{203}+(-42s_1-168)\sqrt{29})\sqrt{2}r_2\displaybreak[0]\nonumber\\
        &+(((40s_1^2+26s_1-788)\sqrt{29}+(232s_1^2-58s_1-4988))\sqrt{7}+((60s_1^2+42s_1-252)\sqrt{29}\displaybreak[0]\nonumber\\
        &+(116s_1^2+406s_1+812))))\sqrt{-1}+((4s_1^2+2s_1-56)\sqrt{203}+(42s_1+168)\sqrt{29})\sqrt{2}r_2\displaybreak[0]\nonumber\\
        &+((-40s_1^2-26s_1+788)\sqrt{29}-232s_1^2+58s_1+4988)\sqrt{7}\displaybreak[0]\nonumber\\
        &+(60s_1^2+42s_1-252)\sqrt{29}+116s_1^2+406s_1+812\displaybreak[3]\nonumber\\[1ex plus 3ex]
x_{20}={}&((18s_1\sqrt{203}+(-42s_1+84)\sqrt{29})\sqrt{2}r_2+(((16s_1^2-34s_1-248)\sqrt{29}-174s_1-696)\sqrt{7}\displaybreak[0]\nonumber\\
        &+((-8s_1^2-322s_1-476)\sqrt{29}-232s_1^2-2030s_1+812)))\sqrt{-1}\displaybreak[0]\nonumber\\
        &+(18s_1\sqrt{203}+(42s_1-84)\sqrt{29})\sqrt{2}r_2+((-16s_1^2+34s_1+248)\sqrt{29}+(174s_1+696))\sqrt{7}\displaybreak[0]\nonumber\\
        &+(-8s_1^2-322s_1-476)\sqrt{29}-232s_1^2-2030s_1+812\displaybreak[3]\nonumber\\[1ex plus 3ex]
x_{21}={}&(((-4s_2^2-4s_2+56)s_1^2+(-8s_2^2+28s_2+112)s_1+(104s_2^2+140s_2-1456))\sqrt{29}\displaybreak[0]\nonumber\\
        &+((36s_2^2-4s_2-336)s_1^2+(216s_2^2-140s_2-2016)s_1-792s_2^2-28s_2+7392))r_1\sqrt{-1}\displaybreak[0]\nonumber\\
        &+((1/7(-4s_2^2-28s_2+56)s_1^2+(-8s_2^2-20s_2+112)s_1+(8s_2^2+92s_2-112))\sqrt{29}\displaybreak[0]\nonumber\\
        &+(1/7(-60s_2^2+84s_2+560)s_1^2+(24s_2^2+36s_2-224)s_1+(264s_2^2-300s_2-2464)))\sqrt{7}r_1\displaybreak[3]\nonumber\\[1ex plus 3ex]
x_{22}={}&((((s_2+2)s_1^2+(2s_2+4)s_1-8s_2-70)\sqrt{29}\displaybreak[0]\nonumber\\
        &+((2s_2^2+3s_2-38)s_1^2+(4s_2^2+6s_2-76)s_1-16s_2^2-24s_2+826))\sqrt{14}r_2\displaybreak[0]\nonumber\\
        &+(((2s_2^2+2s_2)s_1^2+(-20s_2^2-56s_2+168)s_1-76s_2^2-112s_2+672)\sqrt{29}\displaybreak[0]\nonumber\\
        &+((-18s_2^2+2s_2+168)s_1^2+(36s_2^2+112s_2-336)s_1+(540s_2^2+56s_2-5040))))r_1\sqrt{-1}\displaybreak[0]\nonumber\\
        &+(((-s_2s_1^2+4s_2s_1+14s_2)\sqrt{29}\displaybreak[0]\nonumber\\
        &+((6s_2^2+9s_2-56)s_1^2+(-24s_2^2-36s_2+224)s_1-84s_2^2-126s_2+784))\sqrt{2}r_2\displaybreak[0]\nonumber\\
        &+((1/7(-2s_2^2-14s_2+56)s_1^2+(8s_2^2+20s_2-56)s_1+(16s_2^2+76s_2-280))\sqrt{29}\displaybreak[0]\nonumber\\
        &+(1/7(-30s_2^2+42s_2+280)s_1^2+(-24s_2^2-36s_2+224)s_1+(96s_2^2-204s_2-896)))\sqrt{7})r_1\displaybreak[3]\nonumber\\[1ex plus 3ex]
x_{23}={}&(((-2s_1^2+(-3s_2-1)s_1+(6s_2+28))\sqrt{29}\displaybreak[0]\nonumber\\
        &+((4s_2^2+6s_2-18)s_1^2+(2s_2^2+3s_2-9)s_1-56s_2^2-84s_2+252))\sqrt{14}r_2\displaybreak[0]\nonumber\\
        &+(((-4s_2^2-10s_2+56)s_1^2+(-2s_2^2-14s_2+28)s_1+(20s_2^2+14s_2-280))\sqrt{29}\displaybreak[0]\nonumber\\
        &+((12s_2^2+18s_2-112)s_1^2+(-30s_2^2+42s_2+280)s_1-204s_2^2+42s_2+1904)))r_1\sqrt{-1}\displaybreak[0]\nonumber\\
        &+(((-2s_2s_1^2+(-s_2-21)s_1+(28s_2-84))\sqrt{29}\displaybreak[0]\nonumber\\
        &+((6s_2^2+9s_2+147)s_1-84s_2^2-126s_2+1596))\sqrt{2}r_2\displaybreak[0]\nonumber\\
        &+((1/7(-8s_2^2-14s_2)s_1^2+(-10s_2^2-22s_2+84)s_1+(40s_2^2+106s_2-168))\sqrt{29}\displaybreak[0]\nonumber\\
        &+(1/7(48s_2^2+14s_2-448)s_1^2+(42s_2^2+34s_2-392)s_1-96s_2^2-202s_2+896))\sqrt{7})r_1\displaybreak[3]\nonumber\\[1ex plus 3ex]
x_{24}={}&(((-9s_1+9s_2)\sqrt{29}+(87s_1-18s_2^2-27s_2+168))\sqrt{14}r_2\displaybreak[0]\nonumber\\
        &+((-84s_1-90s_2^2-252s_2+672)\sqrt{29}+(162s_2^2+504s_2-1512)))r_1\sqrt{-1}\displaybreak[0]\nonumber\\
        &+(((2s_2s_1^2+(-2s_2-21)s_1-49s_2+42)\sqrt{29}\displaybreak[0]\nonumber\\
        &+((12s_2^2+18s_2-112)s_1^2+(12s_2^2+18s_2+91)s_1-126s_2^2-189s_2+770))\sqrt{2}r_2\displaybreak[0]\nonumber\\
        &+((1/7(20s_2^2+56s_2-224)s_1^2+(4s_2^2+4s_2-28)s_1-58s_2^2-148s_2+616)\sqrt{29}\displaybreak[0]\nonumber\\
        &+(1/7(-36s_2^2-112s_2+336)s_1^2+(-36s_2^2+4s_2+336)s_1+(162s_2^2+272s_2-1512)))\sqrt{7})r_1\displaybreak[3]\nonumber\\[1ex plus 3ex]
x_{25}={}&(((-9s_1+9s_2)\sqrt{29}+(87s_1-18s_2^2-27s_2+168))\sqrt{14}r_2\displaybreak[0]\nonumber\\
        &+((84s_1+(90s_2^2+252s_2-672))\sqrt{29}+(-162s_2^2-504s_2+1512)))r_1\sqrt{-1}\displaybreak[0]\nonumber\\
        &+(((2s_2s_1^2+(-2s_2-21)s_1-49s_2+42)\sqrt{29}\displaybreak[0]\nonumber\\
        &+((12s_2^2+18s_2-112)s_1^2+(12s_2^2+18s_2+91)s_1-126s_2^2-189s_2+770))\sqrt{2}r_2\displaybreak[0]\nonumber\\
        &+((1/7(-20s_2^2-56s_2+224)s_1^2+(-4s_2^2-4s_2+28)s_1+(58s_2^2+148s_2-616))\sqrt{29}\displaybreak[0]\nonumber\\
        &+(1/7(36s_2^2+112s_2-336)s_1^2+(36s_2^2-4s_2-336)s_1-162s_2^2-272s_2+1512))\sqrt{7})r_1\displaybreak[3]\nonumber\\[1ex plus 3ex]
x_{26}={}&(((-2s_1^2+(-3s_2-1)s_1+(6s_2+28))\sqrt{29}\displaybreak[0]\nonumber\\
        &+((4s_2^2+6s_2-18)s_1^2+(2s_2^2+3s_2-9)s_1-56s_2^2-84s_2+252))\sqrt{14}r_2\displaybreak[0]\nonumber\\
        &+(((4s_2^2+10s_2-56)s_1^2+(2s_2^2+14s_2-28)s_1-20s_2^2-14s_2+280)\sqrt{29}\displaybreak[0]\nonumber\\
        &+((-12s_2^2-18s_2+112)s_1^2+(30s_2^2-42s_2-280)s_1+(204s_2^2-42s_2-1904))))r_1\sqrt{-1}\displaybreak[0]\nonumber\\
        &+(((-2s_2s_1^2+(-s_2-21)s_1+(28s_2-84))\sqrt{29}\displaybreak[0]\nonumber\\
        &+((6s_2^2+9s_2+147)s_1-84s_2^2-126s_2+1596))\sqrt{2}r_2\displaybreak[0]\nonumber\\
        &+((1/7(8s_2^2+14s_2)s_1^2+(10s_2^2+22s_2-84)s_1-40s_2^2-106s_2+168)\sqrt{29}\displaybreak[0]\nonumber\\
        &+(1/7(-48s_2^2-14s_2+448)s_1^2+(-42s_2^2-34s_2+392)s_1+(96s_2^2+202s_2-896)))\sqrt{7})r_1\displaybreak[3]\nonumber\\[1ex plus 3ex]
x_{27}={}&((((s_2+2)s_1^2+(2s_2+4)s_1-8s_2-70)\sqrt{29}\displaybreak[0]\nonumber\\
        &+((2s_2^2+3s_2-38)s_1^2+(4s_2^2+6s_2-76)s_1-16s_2^2-24s_2+826))\sqrt{14}r_2\displaybreak[0]\nonumber\\
        &+(((-2s_2^2-2s_2)s_1^2+(20s_2^2+56s_2-168)s_1+(76s_2^2+112s_2-672))\sqrt{29}\displaybreak[0]\nonumber\\
        &+((18s_2^2-2s_2-168)s_1^2+(-36s_2^2-112s_2+336)s_1-540s_2^2-56s_2+5040)))r_1\sqrt{-1}\displaybreak[0]\nonumber\\
        &+(((-s_2s_1^2+4s_2s_1+14s_2)\sqrt{29}\displaybreak[0]\nonumber\\
        &+((6s_2^2+9s_2-56)s_1^2+(-24s_2^2-36s_2+224)s_1-84s_2^2-126s_2+784))\sqrt{2}r_2\displaybreak[0]\nonumber\\
        &+((1/7(2s_2^2+14s_2-56)s_1^2+(-8s_2^2-20s_2+56)s_1-16s_2^2-76s_2+280)\sqrt{29}\displaybreak[0]\nonumber\\
        &+(1/7(30s_2^2-42s_2-280)s_1^2+(24s_2^2+36s_2-224)s_1-96s_2^2+204s_2+896))\sqrt{7})r_1\nonumber
\end{align}
\end{footnotesize}%
Finally, the (non-normalized) fiducial vector with respect to the standard basis can be
obtained as $|\widetilde{\psi}\rangle=T|\widetilde{\psi}_0\rangle$.

We have also tried to compute a fiducial vector for dimension
$N=18=3^2\times 2$.  In that case, an even more sparse representation
of the Clifford group with at most two entries per row and column can
be obtained. The problem is that for $N=18$ we only have the Zauner
symmetry (as indicated by the numerical solutions), which means we
have to search for a fiducial vector within a $7$-dimensional complex
vector space, resulting in $13$ real parameters.  By contrast for
$N=28$ the additional symmetry reduces the number of real parameters
to only $10$.  It turns out that the disadvantage, of having more
parameters for $N=18$, outweighs the advantage, of a more sparse group
representation.  Consequently we were unable to obtain a solution in
the time available.

\section{Conclusion}
\label{sec:conclusion}
In this paper we showed that the standard representation of the Clifford group has a system of imprimitivity consisting of $k$-dimensional subspaces, where $k$ is the square free part of the dimension.  This means that one can choose a basis such that the representation matrices are all what we call $k$-nomial, with exactly one non-zero $k\times k$ block in each row and each column of blocks.  We then used this basis to perform a hand-calculation of the
(already known) exact fiducials in dimension $8$, and to obtain by
machine-calculation a new exact fiducial in dimension $28$.  We also revisited
the machine calculation of a fiducial in dimension $12$ having only the Zauner
symmetry, and found that the computation time was reduced by a factor of more
than $10^3$, and that the solution obtained was much more compact.  Our results suggest that the $k$-nomial basis is better adapted to
the SIC-problem than the standard basis.  As we remarked in the Introduction,
the fact that SICs exist at all (analytically or numerically) in every dimension
$\le 67$ in spite of the defining equations being greatly over-determined, and
the fact that the Galois group of the known exact fiducials has a particularly
simple form even among the class of solvable groups, both hint at the presence
of some underlying symmetry or other algebraic feature of the equations which
has hitherto escaped notice.  Our hope is that the work reported here will help
us to discover that hidden feature and so lead to a solution to the
SIC-existence problem.

The Clifford group has numerous applications.  It appears to us that
the sparsity of the representation matrices described here means that
they are likely to be relevant to other problems, apart from the
SIC-existence problem.

As we mentioned in the Introduction there are several different,
though closely related constructions which go by the name ``Clifford
group''.  Besides the version of the group considered here there is
the version defined to be the normalizer of the tensor product of
several copies of the Weyl-Heisenberg group~\cite{Hostens,Gross}, and
there is also what might be called the Galoisian version defined in
prime power dimension using a finite field~\cite{Vourdas,Appleby09a}.
It would be interesting to try to extend the analysis of this paper to
this more general setting.

\section*{Acknowledgements}
We thank Jon Yard for valuable discussions. 
DMA was supported in part
by the U.~S. Office of Naval Research (Grant No.\ N00014-09-1-0247)
and by the John Templeton Foundation.  IB was supported by the Swedish Research Council 
under contract VR 621-2010-4060. \AA.E. acknowledges support  from the Wenner-Gren foundations, and hospitality and support from  Institut Mittag-Leffler where part of the research for this paper was carried out.   Research at Perimeter Institute
is supported by the Government of Canada through Industry Canada and
by the Province of Ontario through the Ministry of Research \&
Innovation.  The Centre for Quantum Technologies is a Research Centre of Excellence
funded by the Ministry of Education and the National Research
Foundation of Singapore.

\end{allowdisplaybreaks}

\end{document}